\documentclass[fleqn,usenatbib]{mnras}

\usepackage{newtxtext,newtxmath}

\usepackage[T1]{fontenc}

\DeclareRobustCommand{\VAN}[3]{#2}
\let\VANthebibliography\thebibliography
\def\thebibliography{\DeclareRobustCommand{\VAN}[3]{##3}\VANthebibliography}

\usepackage{graphicx}	
\usepackage{amsmath}	
\usepackage[normalem]{ulem}
\usepackage{multirow}
\usepackage[table,xcdraw]{xcolor}

\title[X-ray polarimetry of distant AGN components]{X-ray polarization from parsec-scale components of active galactic nuclei: observational prospects}

\author[J. Podgorn\'y et al.]{
J. Podgorn\'y$^{1,2,3}$\thanks{E-mail: jakub.podgorny@asu.cas.cz},
F. Marin$^{1}$ 
and M. Dov{\v{c}}iak$^{2}$ 
\\
$^{1}$Universit\'e de Strasbourg, CNRS, Observatoire Astronomique de Strasbourg, UMR 7550, F-67000 Strasbourg, France\\
$^{2}$Astronomical Institute, Academy of Sciences of the Czech Republic, Bo{\v{c}}n\'i II, CZ-14131 Prague, Czech Republic\\
$^{3}$Astronomical Institute, Charles University, V Hole{\v{s}}ovi{\v{c}}k\'ach 2, CZ-18000 Prague, Czech Republic\\
}

\date{Accepted XXX. Received YYY; in original form ZZZ}

\pubyear{2023}

\begin{document}
\label{firstpage}
\pagerange{\pageref{firstpage}--\pageref{lastpage}}
\maketitle

\begin{abstract}

We present a broad analysis of X-ray polarimetric observational prospects for radio-quiet active galactic nuclei (AGN), focusing on the role of parsec-scale components. We provide a revision of self-consistent type-1 and type-2 generic AGN radiative transfer models that were obtained with a Monte Carlo code {\tt STOKES}, evaluating the effects of absorption and scattering. Our model consists of a central disc-corona emission obtained with the {\tt KYNSTOKES} code in the lamp-post geometry, an equatorial wedge-shaped dusty torus and two symmetric conical polar outflows. We argue that the information on the mutual orientation, shape, relative size and composition of such components, usually obtained from spectroscopy or polarimetry in other wavelengths, is essential for the X-ray polarization analysis of the obscured type-2 AGNs. We provide general detectability prospects for AGNs with 2--8 keV polarimeters on board of the currently flying IXPE satellite and the forthcoming eXTP mission. Finally, we assess the role of contemporary X-ray polarimetry in our understandings of the unified AGN model after the first year and a half of IXPE operation.

\end{abstract}




\section{Introduction}\label{introduction}

Active galactic nuclei (AGN) are one of the intrinsically brightest known objects on the sky at all wavelengths. They are formed when matter is accreted in a form of a disc onto a supermassive black hole, which is present in nearly every massive galaxy, whether it is actively accreting or not \citep[see, e.g.][]{Pringle1972, Shakura1973, Seward2010}. The observer's inclination towards the axially symmetric structure, surrounded by a parsec-scale dusty torus in the equatorial plane, determines the spectral and polarization properties of the source. We therefore classify AGNs as type-1 (pole-on view) and type-2 (edge-on view), depending of the line of sight of the observer \citep{Rowan-Robinson1977, Keel1980, Antonucci1993}. When the black hole spins and the accretion disc is strongly magnetized, powerful relativistic jets are formed, ejecting material from the vicinity of the central black hole up to megaparsec in highly colimated (spanning only a few angular degrees) directions from the poles \citep[for a recent review, see e.g.][]{Blandford2019}. In this study, we will focus on those AGNs that are not viewed directly through the jet directions (labelled as blazars) and that do not exhibit strong jets (labelled as radio-loud AGNs).

This unification scenario \citep{Antonucci1993} can be effectively examined in X-rays, including polarimetry in addition to the standard spectroscopic and timing observational techniques. The Imaging X-ray Polarimetry Explorer (IXPE) \citep{Weisskopf2022}, observing since early 2022, has opened the possibilities of polarization measurements in the 2--8 keV band. A handful of AGNs were observed during the first year and a half of IXPE observations: the type-1 MCG 05-23-16 \citep[$< 3.2\%$ in 2--8 keV at a 99\% confidence level,][]{Marinucci2022, Tagliacozzo2023}, NGC 4151 \citep[$4.9 \pm 1.1$ \% in 2--8 keV at a 68\% confidence level,][]{Gianolli2023}, and IC 4329A \citep[$3.3 \pm 1.9$ \% in 2--8 keV at a 90\% confidence level,][]{Ingram2023}; and one type-2 AGN, the Circinus Galaxy \citep[$20.0 \pm 3.8$ \% in 2--6 keV at a 68\% confidence level,][]{Ursini2023}.

Naturally, alongside the recent advances in observations, theoretical X-ray polarization models of AGNs are significantly improving. First attempts to produce a global self-consistent X-ray polarization model of AGN were done by \cite{Goosmann2011, Marin2012b, Marin2013, Marin2016b}, focused on the sources NGC 1068, MCG 06-30-15, NGC 1365 and NGC 4151, respectively. All of these used the Monte Carlo radiative transfer code {\tt STOKES} \citep{Goosmann2007,Marin2012,Marin2015,Marin2018} for their predictions. In particular, \cite{Goosmann2011, Marin2016b} added a model of a homogenous polar scatterer to the dusty torus and an additional equatorial scattering ring located in the transition region between the parsec-scale dusty torus and the accretion disc \citep{Antonucci1984, Smith2004}. More recently, \cite{Marin2018b, Marin2018c} \citep[and in application to eclipsing events in][]{Kammoun2018} made a broad theoretical X-ray polarimetric study of type-2 and type-1 AGNs, respectively, without any focus on a particular AGN. Using the {\tt STOKES} code, they kept a homogenous polar component self-consistently next to a dusty torus of uniform density, but without the additional equatorial scattering ring. Examining a large parametric space step by step, they have shown that distant reprocessing in both of the components has a non-negligible impact on the total X-ray polarization outcome for both type-1 and type-2 AGNs. In the paper \cite{Podgorny2023b} (hereafter Paper I.), we have recently presented an updated and revised summary of the X-ray spectro-polarimetric properties of the equatorial parsec-scale AGN components (i.e. the dusty tori), using the same method. Having studied the sensitivity of the reprocessed X-ray polarization outcome to a particular toroidal structure in Paper I. in detail, we argued that the results presented in \cite{Marin2018b, Marin2018c} form only a first step in the exploration of the configuration space that our current knowledge of AGNs enables, and that also the realistic total X-ray polarization output can be more diverse even for media with the simplified assumption of uniform density.

Analyzing each AGN component separately can give useful insights on the decomposition of the total X-ray polarization, knowing the relative flux contributions from spectroscopy. However, this decomposition is often limited in practice, because two components polarized in orthogonal directions cancel each other out in terms of superposition of polarization vectors and the degeneracies are further complicated by interactions of the components (through the exchange of scattered photons and by mutual dynamical interaction). In this paper, we will follow up by studying one type of equatorial scattering geometry from Paper I., the wedge-shaped torus, and by incorporating it into a toy model of a full AGN. Using our latest X-ray polarization model {\tt KYNSTOKES} presented in \cite{Podgorny2023} for the inner-most disc-corona emission in the so-called lamp-post geometry \citep{Matt1991,Martocchia1996,Henri1997,Petrucci1997, Martocchia2000, Miniutti2004, Dovciak2004b, Dovciak2011, Parker2015, Furst2015, Miller2015, Niedzwiecki2016, Dovciak2016, Walton2017, Ursini2020}, we will revise the total AGN model presented in \cite{Marin2018b, Marin2018c} and point out intricacies that have not been studied yet in such scenario and that can affect the predicted X-ray signal for a distant observer. Other self-consistent models of the total X-ray polarimetric output of AGNs, given by the reprocessing in circumnuclear components of more diverse nature described in Paper I., and by adding e.g. the broad-line regions (BLRs), radiation-driven or magneto-hydro-dynamical wind models, or incorporating different coronal geometries, are left for future investigation. In this paper we will rather argue, providing a few examples in a simple, axially symmetric and static 3-component model [the central lamp-post disc-corona emission, the homogenous equatorial dusty torus, and the homogenous polar outflows representing the narrow-line regions (NLRs)], that we are still far from lifting the X-ray polarization degeneracies in AGNs and far from having an efficient approximative tool to give sharp observational constraints on the outer geometry and composition of accreting supermassive black holes \textit{without a focus on a particular source} due to the complexity of AGNs. We will at least attempt to suggest particular configurations, where energy-resolved X-ray polarization observations are more likely to bring any insight. We will omit the discussion of time-resolved X-ray polarimetry, which is not of primary interest for AGNs, given the sensitivy of the current and forthcoming X-ray polarimeters.

Last but not least, the aim of this work is to enclose the link between our latest modeling efforts and observations through simulating the observations of our modelled AGNs by IXPE that will continue bringing results in this decade. Doing so, we may estimate \textit{general} detectability prospects of AGNs in the mid X-rays, which is useful before any more ambitious attempt to characterize and quantify the amount of degeneracies in the description of AGN components that current 2--8 keV X-ray polarimetry can lift from spectroscopy, timing analysis, or polarimetry in other wavelengths. The future X-ray polarimetric targeting of AGNs and observational planning can benefit from such guidelines, simultaneously with the detailed analysis of \textit{particular} sources that is already available after the first year and a half of IXPE operations. All of this is timely, because the IXPE observations of type-1 and type-2 radio-quiet AGNs will soon form a larger sample, from which we may be able to better assess the AGN unification scenario.

The paper is organized as follows: in Section \ref{methods} we introduce the assumptions and numerical models. Section \ref{fullmodel} provides the X-ray spectro-polarimetric modelling results for one type of self-consistent AGN models by revisiting the published computations in \cite{Marin2018b, Marin2018c}. The resulting up-to-date observational prospects are given in Section \ref{observations}. We conclude in Section \ref{conclusions}.

\section{Definitions and methods}\label{methods}

We define the linear polarization degree $p$ and linear polarization angle $\Psi$ in the usual way from the Stokes parameters $I$, $Q$ and $U$:
\begin{equation}\label{ppsidef}
	\begin{aligned}
		p &= \dfrac{\sqrt{Q^2+U^2}}{I} \\
		\Psi &= \dfrac{1}{2}\textrm{\space}\arctan_2\left(\dfrac{U}{Q}\right)   \textrm{ ,}
	\end{aligned}
\end{equation}
where $\arctan_2$ denotes the quadrant-preserving inverse of a tangent function and $\Psi = 0$ means that the polarization vector is oriented \textit{parallel} to the system axis of symmetry projected to the polarization plane. $\Psi$ increases in the counter-clockwise direction from the point of view of an incoming photon. We will use the notation $p_0$ for the primary polarization fraction assigned to the isotropic coronal source of emission in the lamp-post geometry (see below for the central emission model implementation, and see Sections \ref{fullmodel} and \ref{ixpekynstokes} for the discussion of other coronal geometries in the context of recent IXPE discoveries).

\begin{figure}
	\includegraphics[width=1.\columnwidth]{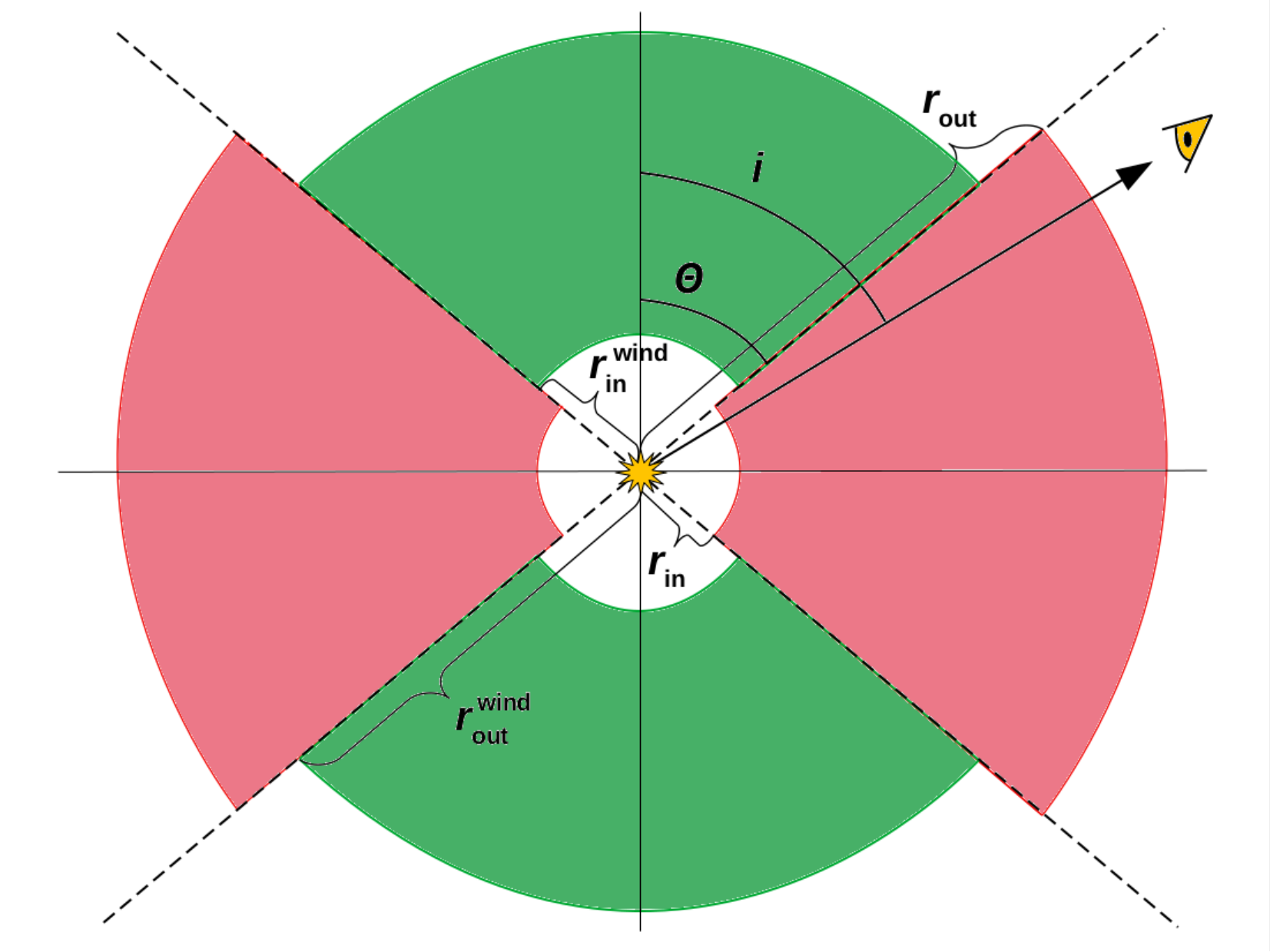}
	\caption{A schematic in the meridional plane of the axially symmetric AGN parsec-scale components probed by {\tt STOKES}. The illuminating disc-corona region is located in the center of the coordinate system and is assumed to be in the lamp-post geometry. We assume a wedge-shaped equatorial scatterer (i.e. Case A of Paper I.), displayed in red and representing the dusty torus, in combination with a cone-shaped polar scatterers, displayed in the green and representing the NLR. We define a type-1 observer, if $i<\Theta$, and type-2 observer, if $i>\Theta$.}
	\label{stokes_mo}
\end{figure}

For the Monte Carlo parsec-scale computations made with {\tt STOKES}\footnote{\url{www.stokes-program.info}} \citep{Goosmann2007,Marin2012,Marin2015,Marin2018} we used the version \textit{v2.07} that is suitable for X-rays and that was also used in \cite{Marin2018b, Marin2018c} and Paper I. The code is appropriate for obtaining the polarization properties of radiation in media where scattering and absorption is the dominant source of opacity. In this paper, we will adopt the same the same 3-component setup as in \cite{Marin2018b, Marin2018c}, where spectro-polarimetric properties of AGNs for type-1 and type-2 viewing angles were discussed for various compositions of equatorial and polar scattering regions. We refer the reader to \cite{Marin2018b, Marin2018c} and Paper I. for all the details on physical processes and the simulation setup that we will use. Paper I. assumed only the equatorial scattering region and a more simple source of central isotropic power-law emission (a 2-component model), but focused in more detail on the reprocessing from equatorial tori of different geometries and content. Comparing to Paper I., we only add to the wedge-shaped torus two identical homogenous polar outflows, representing the NLRs, in the same wedge-like geometry (rotated 90\degr in the meridional plane and being axially symmetric around the principal system axis). See Figure \ref{stokes_mo} for a sketch of the model with the geometrical parameters indicated. We refer to \cite{Marin2018b, Marin2018c} for the choice of parametric values that will be used in Section \ref{fullmodel}, if not stated otherwise. The conical shape of the polar scatterers is determined by the inner and outer radii $r_\textrm{in}^\textrm{wind}$ and $r_\textrm{out}^\textrm{wind}$, similarly to the inner and outer radii of the equatorial region $r_\textrm{in}$ and $r_\textrm{out}$.

The composition of the equatorial and polar scatterers assumes the same atomic species used for \cite{Marin2018b, Marin2018c} and the solar abundance from \cite{Asplund2005} with $A_\mathrm{Fe} = 1.0$. For the equatorial ``tori'', the uniform neutral hydrogen density $n_\textrm{H}$ is given by the total column density $N_\textrm{H} = n_\textrm{H}L$, where $L = r_\textrm{out} - r_\textrm{in}$ is the size of the scattering region between the center and any type-2 observer (for the wedge-like geometry the line-of-sight size of the region is not dependent on inclination). The same applies to the conical polar outflows and any imaginary type-1 observer, we will just use a different notation for the total column density $N_\textrm{H}^\textrm{wind} = n_\textrm{H}^\textrm{wind}L^\textrm{wind}$, where $L^\textrm{wind} = r_\textrm{out}^\textrm{wind} - r_\textrm{in}^\textrm{wind}$. We also define for both types of regions separately the optical depth for electron scattering $\tau_\textrm{e} = \sigma_\textrm{T} n_\textrm{e} L$ and $\tau_\textrm{e}^\textrm{wind} = \sigma_\textrm{T} n_\textrm{e}^\textrm{wind} L^\textrm{wind}$ via the Thomson cross section $\sigma_\textrm{T}$ and homogenous free electron density $n_\textrm{e}$ and $n_\textrm{e}^\textrm{wind}$, defining the level of ionization in the medium. We opted for $\tau_\textrm{e}^\textrm{wind} = 0.03$ in the cases of ionized polar winds that were studied among others for type-2 AGNs. Anything above this free electron optical depth showed overly ionized winds, which then determined the total emission entirely and the rest of the parametric space became uninteresting.

We replaced the central emission model used in \cite{Marin2018b, Marin2018c} by a more complex one in this study.\footnote{The original studies included the lamp-post disc-corona emission described in \cite{Dovciak2011} with Chandrasekhar's approximation for disc reprocessing \citep{Chandrasekhar1960}, neutral disc only and incorrect computations of the relativistic change of the polarization angle between the lamp and the disc, which is now fixed.} The central source is provided by the output of the latest version of the spectro-polarimetric {\tt KYNSTOKES} code \citep{Podgorny2023}, assuming unpolarized and $2\%$ polarized emission (parallely and perpendicularly to the axis of symmetry) at the location of lamp-post corona that is isotropically illuminating towards the disc and the observer. The code allows to estimate X-ray spectro-polarimetric output from the corona alongside relativistic reflection from a geometrically thin and optically thick disc, all for a distant observer located at arbitrary inclination with respect to the disc. {\tt KYNSTOKES} includes all general-relativistic (GR) effects, apart from returning radiation (i.e. secondary reflections from the disc), and assumes an X-ray coronal power-law at the lamp-post location on the rotation axis of a black hole with spin $a$ in a Kerr metric. The accretion disc located in the equatorial plane is extending from the inner-most circular orbit (ISCO) to 400 gravitational radii from the black hole -- a value above which the outer extension of the disc no longer impacts the results of the model \citep{Podgorny2023}. The disc is newly allowed to be partially ionized, therefore in this study we test a) the almost fully ionized disc ($L_\textrm{X}/L_\textrm{Edd} = 0.1 $ and $M_\textrm{BH}/M_\odot = 10^5 $ in {\tt KYNSTOKES}) vs. b) the almost fully neutral disc case \cite[$L_\textrm{X}/L_\textrm{Edd} =  0.001$ and $M_\textrm{BH}/M_\odot =  10^8 $  in {\tt KYNSTOKES}, comparable to the neutral disc computations in][]{Marin2018b,Marin2018c} to test the effects of disc ionization in a global view. We refer to \cite{Podgorny2023} for all the information on the {\tt KYNSTOKES} model and to \cite{Marin2018b, Marin2018c} for the incorporation of such central X-ray emission into the ``full'' AGN model, which we preserved in Section \ref{fullmodel}, i.e. we illuminate each scattering region isotropically with the {\tt KYNSTOKES} output for inclinations $i =20\degr$ and $i = 70\degr$ (measured from the pole) towards the polar and towards the equatorial scatterers, respectively.

The revision and one-to-one comparisons with previous studies that were carried with the lamp-post model are the main reason for its usage here. The explored parameter space is already large and simulating another disc-corona geometry would be of scope of another paper. We will provide elementary predictions for coronae elongated in the equatorial plane in the next sections, although they were not simulated within our 3-component model. As first-order estimates, the different orientation of the primary polarization may serve for results on various geometries, because for lamp-posts, due to the location of the disc origin of seed photons with respect to the Comptonizing medium, we assume rather perpendicular polarization to the principal axis, while the opposite is typically the theoretical result for slab coronal geometries, elongated in the equatorial plane \citep{Ursini2022, Krawczynski2022, Krawczynski2022b}.

Also the choice of unpolarized and 2\% polarized primary radiation is well within conservative estimates for Comptonization of disc seed photons. We note that the recent IXPE observations suggest in some AGNs or X-ray binary systems (XRBs) even about twice higher coronal polarization \citep{Krawczynski2022, Gianolli2023, Ingram2023, Dovciak2023}, but the aim of this paper is to assess a generic sample and the sensitivity to changing primary polarization, given by three different incident polarization states. The choice of unpolarized primary and $2\%$ polarized primary in the two orthogonal directions was also adopted in the previous studies to which we will directly compare.

Tables \ref{table_model_grid_type1} and \ref{table_model_grid_type2} summarize the main simulation cases that were examined with respect to \cite{Marin2018b,Marin2018c}. The values correspond to the average 2--8 keV model polarization that will be elaborated on in Section \ref{fullmodel} and to the detectability prospects that will be in detail examined in Section \ref{observations}.
\begin{table}
\centering
	\caption{Summary of the main configuration space in this study for type-1 AGNs. In addition to \citet{Marin2018b,Marin2018c} the possibility of partial disc ionization was tested with {\tt KYNSTOKES}. The column densities $N_\textrm{H}$ are given in $\textrm{cm}^{-2}$. We assume the notation of positive or negative polarization, if the corresponding polarization angle is parallel or perpendicular to the axis of symmetry, respectively. Top value in each slot (black bold) is the model unfolded 2--8 keV average polarization degree, $p$, for a given configuration. The three values in left column in each slot (blue) are the estimated observational times in Ms, $T_\textrm{obs}$, that are needed for the model $|p|$ to exceed the simulated $M\!D\!P$ for IXPE (evaluates whether the polarization is to be detected at a 99\% confidence level, although here we compare only the unfolded model), using the unweighted approach in {\tt IXPEOBSSIM}, if the observed X-ray flux is $F_\mathrm{X, 2-10} = 1 \times 10^{-10}, 5 \times 10^{-11}, 1 \times 10^{-11} \, \textrm{erg cm}^{-2} \, \textrm{s}^{-1}$ from top to bottom, respectively, and that were linearly interpolated in the computed $\{T_\textrm{obs},F_\textrm{X,2-10}\}$ space. The three values in right column in each slot (red) are the estimated observed X-ray fluxes in $10^{-10} \, \textrm{erg cm}^{-2} \, \textrm{s}^{-1}$, $F_\mathrm{X, 2-10}$, that are needed for the model $|p|$ to exceed the simulated $M\!D\!P$, using the unweighted approach in {\tt IXPEOBSSIM}, if the observational time is $T_\mathrm{obs} = 0.5, 1, 1.5 \, \textrm{Ms}$ from top to bottom, respectively, and that were linearly interpolated in the computed $\{T_\textrm{obs},F_\textrm{X,2-10}\}$ space. The results are serving as first-order estimates only, see text for details. For a particular source, more specific information on its unresolved composition may be available, which allows detectability predictions with higher accuracy. The $M\!D\!P$ values can be also slightly reduced by means of weighted approach in {\tt IXPEOBSSIM}, which is available for real data analysis. The observational times needed for the IXPE mission may reduce by a factor of 4 for the planned eXTP mission due to its larger effective mirror area.}
 \resizebox{\columnwidth}{!}{%
\begin{tabular}{ccc|llllll|ll|ll|}
\cline{4-13}
\multicolumn{1}{l}{}            & \multicolumn{1}{l}{}    & \multicolumn{1}{l}{} & \multicolumn{10}{c}{{\color[HTML]{000000} type-1 AGNs (absorbing winds only)}}                                                                                                                                                                                                                                                                                                                                                                                                                            \\ \cline{4-13} 
\multicolumn{1}{l}{}            & \multicolumn{1}{l}{}    & \multicolumn{1}{l}{} & \multicolumn{6}{c}{{\color[HTML]{000000} $N_\textrm{H}^\textrm{wind} =10^{21} $}}                                                                                                                     & \multicolumn{2}{c}{{\color[HTML]{000000} $N_\textrm{H}^\textrm{wind} =10^{22} $}}                                                                                                                     &  \multicolumn{2}{c}{{\color[HTML]{000000}  $N_\textrm{H}^\textrm{wind} =10^{23}$}}                                                                                                                                                                                   \\ \cline{4-13} 
\multicolumn{1}{l}{}            & \multicolumn{1}{l}{}    & \multicolumn{1}{l}{} & \multicolumn{2}{l}{{\color[HTML]{000000} $N_\textrm{H}=10^{23}$}}       & \multicolumn{2}{l}{{\color[HTML]{000000} $N_\textrm{H}=10^{24}$}}       & \multicolumn{2}{l}{{\color[HTML]{000000} $N_\textrm{H}=10^{25} $}}       & \multicolumn{2}{l}{{\color[HTML]{000000} $N_\textrm{H}=10^{24} $}}       & \multicolumn{2}{l}{{\color[HTML]{000000} $N_\textrm{H}=10^{24}$}}           \\
\hline \hline
\end{tabular}}
\label{table_model_grid_type1}
\end{table}

\begin{table*}
\centering
	\caption{The same as in Table \ref{table_model_grid_type1}, but for the simulated type-2 AGNs. The tested fluxes for interpolation of observational times (blue values from top to bottom in each configuration) are $F_\mathrm{X, 2-10} = 1 \times 10^{-11}, 9 \times 10^{-12}, 8 \times 10^{-12} \, \textrm{erg cm}^{-2} \, \textrm{s}^{-1}$.}
 \resizebox{0.8\textwidth}{!}{%
\begin{tabular}{ccc|llllll|llllll|llllll|}
\cline{4-21}
\multicolumn{1}{l}{}            & \multicolumn{1}{l}{}    & \multicolumn{1}{l}{}                                                                                                        & \multicolumn{18}{c}{{\color[HTML]{000000} type-2 AGNs}}                                                                                                                                                                                                                                                                                                                                                                                                                                                           \\ \cline{4-21} 
\multicolumn{1}{l}{}            & \multicolumn{1}{l}{}    & \multicolumn{1}{l}{}                                                                                   & \multicolumn{6}{c}{{\color[HTML]{000000} absorbing winds with $N_\textrm{H}^\textrm{wind} =10^{21}$}}                                                                                                                     & \multicolumn{6}{c}{{\color[HTML]{000000} ionized winds}}                                                                                                                     & \multicolumn{6}{c}{{\color[HTML]{000000} no winds}}                                                                                                                     \\
\hline \hline
\end{tabular}}
\label{table_model_grid_type2}
\end{table*}

\section{Revision of the parsec-scale AGN modelling}\label{fullmodel}

\begin{figure}
	\includegraphics[width=1.\columnwidth]{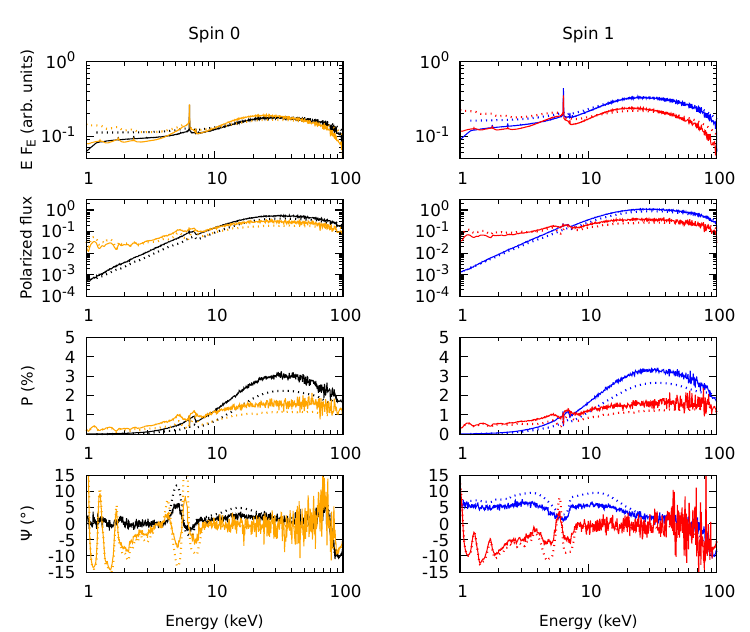}
	\caption{The total AGN emission in the $i = 20\degr$ direction in the case of unpolarized primary radiation (solid lines) and the incident disc-corona emission in the polar direction (dotted lines). We display from top to bottom the energy-dependent flux $EF_\textrm{E}$ (in arbitrary units), the polarized flux, the polarization degree and the polarization angle. Left: the case of black-hole spin 0, right: the case of black-hole spin 1. The computations from \citet{Marin2018c} are displayed in black and blue. The new computations for \textit{ionized} disc are displayed in yellow and red. The winds are neutral with the column density $N_\textrm{H}^\textrm{wind} = 10^{21} \, \textrm{cm}^{-2}$. See \citet{Marin2018c} for the remaining parameters.}
	\label{ionizedKY_UnpolRG_Results_nh24_absorbing_winds}
\end{figure}

\begin{figure}
	\includegraphics[width=1.\columnwidth]{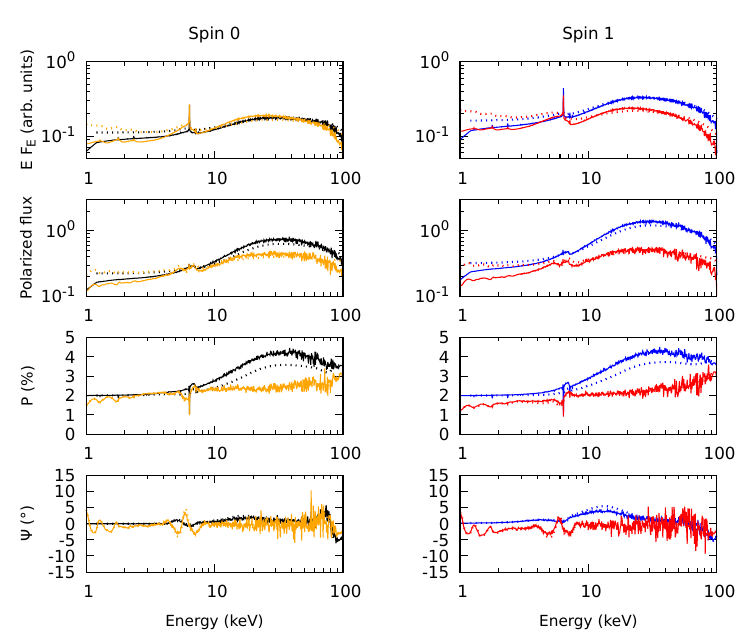}
	\caption{The same as Figure \ref{ionizedKY_UnpolRG_Results_nh24_absorbing_winds}, but for $2\%$ parallelly polarized coronal radiation.}
	\label{ionizedKY_PolGRPara_Results_nh24_absorbing_winds}
\end{figure}

\begin{figure}
	\includegraphics[width=1.\columnwidth]{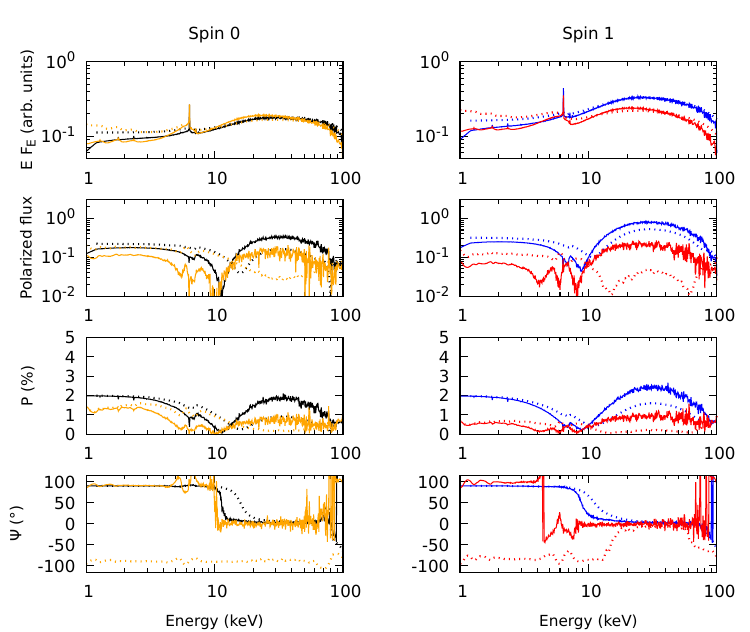}
	\caption{The same as Figure \ref{ionizedKY_UnpolRG_Results_nh24_absorbing_winds}, but for $2\%$ perpendicularly polarized coronal radiation.}
	\label{ionizedKY_PolGRPerp_Results_nh24_absorbing_winds}
\end{figure}

\begin{figure}
	\includegraphics[width=1.\columnwidth]{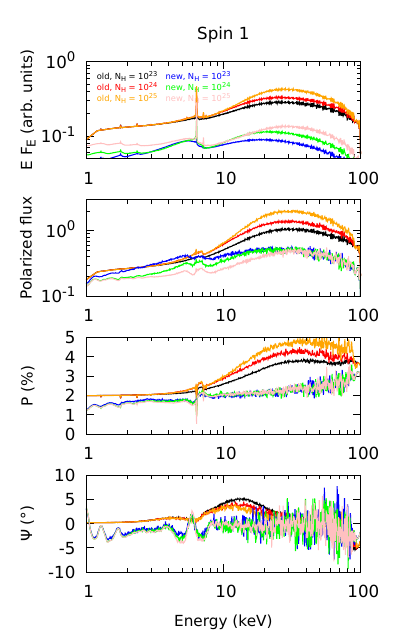}
	\caption{The total AGN emission in the $i = 20\degr$ direction in the case of $2\%$ parallelly polarized primary radiation and black-hole spin 1. We display from top to bottom the energy-dependent flux $EF_\textrm{E}$ (in arbitrary units), the polarized flux, the polarization degree and the polarization angle. We display the results for torus column densities: $N_\textrm{H} = 10^{23} \, \textrm{cm}^{-2}$ (black), $N_\textrm{H} = 10^{24} \, \textrm{cm}^{-2}$ (red), $N_\textrm{H} = 10^{25} \, \textrm{cm}^{-2}$ (orange) for the computations from \citet{Marin2018c} and the same torus densities $N_\textrm{H} = 10^{23} \, \textrm{cm}^{-2}$ (blue), $N_\textrm{H} = 10^{24} \, \textrm{cm}^{-2}$ (green), $N_\textrm{H} = 10^{25} \, \textrm{cm}^{-2}$ (pink) for the new computations for \textit{ionized} disc. The winds are neutral with the column density $N_\textrm{H}^\textrm{wind} = 10^{21} \, \textrm{cm}^{-2}$. See \citet{Marin2018c} for the remaining parameters.}
	\label{ionizedKY_Torus_comparison_para}
\end{figure}

\begin{figure}
	\includegraphics[width=1.\columnwidth]{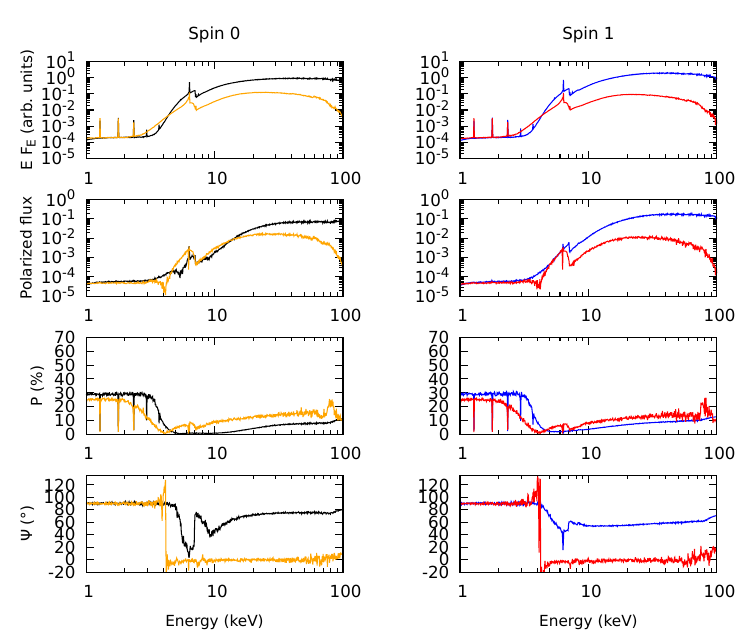}
	\caption{The total AGN emission in the $i = 70\degr$ direction in the case of $2\%$ parallelly polarized coronal radiation. We display from top to bottom the energy-dependent flux $EF_\textrm{E}$ (in arbitrary units), the polarized flux, the polarization degree and the polarization angle. Left: the case of black-hole spin 0, right: the case of black-hole spin 1. The computations from \citet{Marin2018b} are displayed in black and blue. The new computations for \textit{ionized} disc are displayed in yellow and red. The torus column density is set to $N_\textrm{H} = 10^{24} \, \textrm{cm}^{-2}$. The winds are neutral with the column density $N_\textrm{H}^\textrm{wind} = 10^{21} \, \textrm{cm}^{-2}$. See \citet{Marin2018b} for the remaining parameters.}
	\label{type2_ionizedKY_PolGRPara_Results_nh24_absorbing_winds}
\end{figure}

\begin{figure}
	\includegraphics[width=1.\columnwidth]{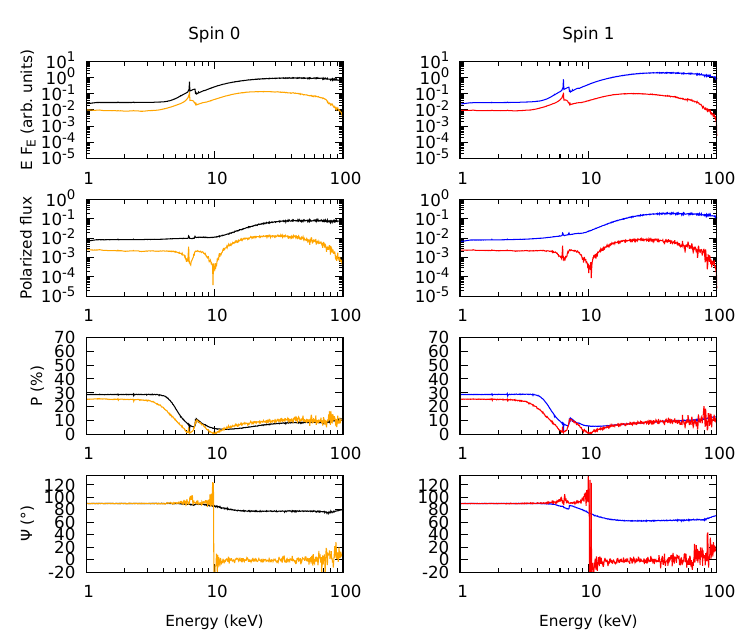}
	\caption{The total AGN emission in the $i = 70\degr$ direction in the case of $2\%$ parallelly polarized coronal radiation. We display from top to bottom the energy-dependent flux $EF_\textrm{E}$ (in arbitrary units), the polarized flux, the polarization degree and the polarization angle. Left: the case of black-hole spin 0, right: the case of black-hole spin 1. The computations from \citet{Marin2018b} are displayed in black and blue. The new computations for \textit{ionized} disc are displayed in yellow and red. The torus column density is set to $N_\textrm{H} = 10^{24} \, \textrm{cm}^{-2}$. The winds are \textit{ionized}. See \citet{Marin2018b} for the remaining parameters.}
	\label{type2_ionizedKY_PolGRPara_Results_nh24_ionized_winds}
\end{figure}

\begin{figure}
	\includegraphics[width=1.\columnwidth]{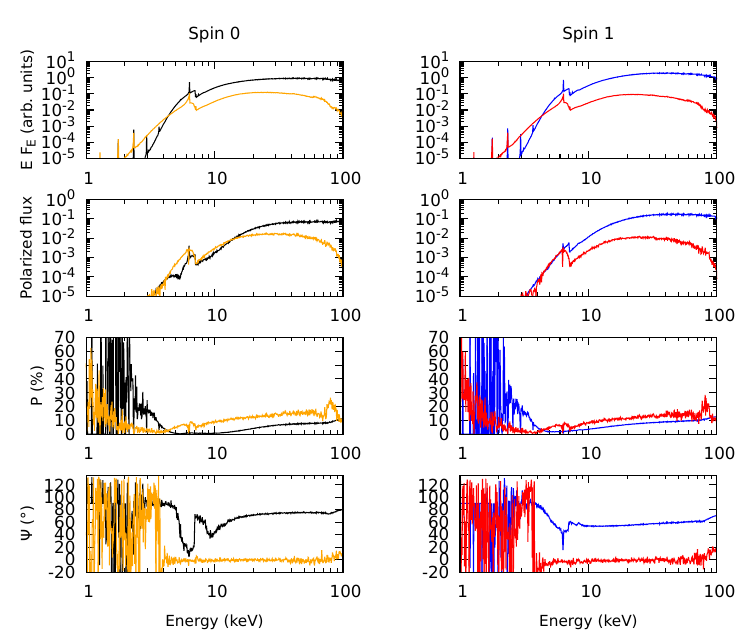}
	\caption{The total AGN emission in the $i = 70\degr$ direction in the case of $2\%$ parallelly polarized coronal radiation. We display from top to bottom the energy-dependent flux $EF_\textrm{E}$ (in arbitrary units), the polarized flux, the polarization degree and the polarization angle. Left: the case of black-hole spin 0, right: the case of black-hole spin 1. The computations from \citet{Marin2018b} are displayed in black and blue. The new computations for \textit{ionized} disc are displayed in yellow and red. The torus column density is set to $N_\textrm{H} = 10^{24} \, \textrm{cm}^{-2}$. There are \textit{no} polar winds. See \citet{Marin2018b} for the remaining parameters.}
	\label{type2_ionizedKY_PolGRPara_Results_nh24_no_winds}
\end{figure}

Although the entire parameter space as in \cite{Marin2018b,Marin2018c} was re-examined, we will plot only the most representative results. We will display the new simulation results in the same way as in \cite{Marin2018b,Marin2018c}, overplotting the former results, so that the reader can notice the changes. We differentiate the old and new computations by a color code. We note that the $\Psi$ orientation in \cite{Marin2018b, Marin2018c} was different from the one stated here by $90\degr$. Thus, for one-to-one comparisons we stick to the definitions stated in our paper and transform the old results accordingly. For clarity purposes, we display all figures comparing the incident radiation in the Appendix \ref{total_comparisons}. Figures \ref{ionizedKY_TF_PO_PA_inputs}--\ref{neutralKY_PERP_TF_PO_PA_inputs_20deg} contain the input results for type-1 AGNs and Figures \ref{type2_ionizedKY_TF_PO_PA_inputs}--\ref{type2_neutralKY_PERP_TF_PO_PA_inputs_70deg} for type-2 AGNs. In the main paper body, we show the total output results for type-1 AGNs in Figures \ref{ionizedKY_UnpolRG_Results_nh24_absorbing_winds}--\ref{ionizedKY_PolGRPerp_Results_nh24_absorbing_winds} for three different incident polarizations, for the case of moderate torus density $N_\textrm{H} = 10^{24} \, \textrm{cm}^{-2}$, for the absorbing neutral winds and for the case of ionized disc, i.e. the disc ionization case that should differ more from the previously published computations. Figure \ref{ionizedKY_Torus_comparison_para} shows the same type-1 configuration with changing torus density for the case of $2\%$ parallely polarized primary. Figures \ref{type2_ionizedKY_PolGRPara_Results_nh24_absorbing_winds}, \ref{type2_ionizedKY_PolGRPara_Results_nh24_ionized_winds} and \ref{type2_ionizedKY_PolGRPara_Results_nh24_no_winds} contain the total output results for type-2 AGNs for the case of moderate torus density $N_\textrm{H} = 10^{24} \, \textrm{cm}^{-2}$, for the ionized disc, for the case of $2\%$ parallely polarized primary, and for absorbing neutral winds, ionized winds, and no polar winds, respectively.

Overall we discovered more discrepancies between our latest results and the previously published results for type-2 AGNs in \cite{Marin2018b}, because of discovered errors in the previous simulation setup and the GR effects described therein, which we no longer see in our latest simulations of type-2 AGNs. Most of the conclusions in the previously published study of type-1 AGNs \cite{Marin2018c} remain valid. In general, we found more dissimilar results in the hard X-ray band than in the 2--8 keV band. Thus most of the conclusions remain valid for the IXPE mission range.

For the type-1s the incident polarization state, including the incident polarization angle, is important for the emerging polarization state change due to circumnuclear reprocessing, which is often neglected in literature when particular unobscured sources are studied. Usually the parallelly polarized input keeps the output polarization degree unaltered, the unpolarized input tends to be additionally polarized by $\approx 1\%$ after the distant reprocessing, while the perpendicularly polarized input tends to be depolarized on the output gradually in energy: from its original $p$ value to zero between 1 and 10 keV. This is because in such configurations the parallel polarization orientation begins to prevail over the perpendicular polarization component above $\sim 10$ keV, which is observed in the polarization angle. This is related to the model of the disc-corona emission (shown in dotted lines for the polar direction of emission) that lacks a strong disc reflection presence at softer energies. There it allows to observe with less dilution the direct primary radiation and its polarization properties, especially for low black-hole spin \citep{Podgorny2023}. For slowly rotating Kerr black holes the co-rotating Novikov-Thorne disc (assumed to extend to ISCO) reaches only six gravitational radii from the black hole \citep{Kerr1963,Novikov1973}. In the equatorial directions the central emission is nearly parallely polarized at all studied X-ray energies (see Figure \ref{type2_ionizedKY_PERP_TF_PO_PA_inputs_70deg}). The reprocessing in distant circumnuclear components adds a contribution of preferred parallel polarization direction at higher X-ray energies, where the photons are also less absorbed in the reprocessed part of the total emission. Compared to \cite{Marin2018c}, the transition in polarization is sharper and occurs at slightly lower energies, which is given by the model differences and the choice of a particular inclination bin. For type-1s, the polarization angle usually remains unaltered from the incident polarization angle, but if the input is unpolarized, it can obtain either parallel or perpendicular polarization, depending on the torus properties and observer's inclination (see Paper I. for more details on this phenomenon). The torus density comparison with energy done in the type-1 paper \cite{Marin2018c} does not agree well with the new simulations, but the reasons remain unknown. It certainly depends on the exact inclination bin chosen in the angular resolution of the Monte Carlo simulation (see below). The wind density comparisons agree. We also observe a slight swing in the polarization angle in the iron line around 6.4 keV, which was not seen in \cite{Marin2018c}. We assume that this is due to lower predicted polarization in this line compared to the previous study, as the polarization angle is undefined for a truly unpolarized fluorescent emission line at its origin.

The Compton-thick type-2 views retain $>25\%$ polarization at soft X-rays (undergoing strong absorption in the torus, with the wind ``periscope'' effect causing the $\Psi = 90\degr$ orientation due to a dominant Compton single-scattering angle in the meridional plane), but low flux with regards to detectability due to significant obscuration. The new GR simulations rather resemble the old ones without GR effects in the type-2 paper \cite{Marin2018b}. After some investigations it was concluded that for this publication the simulations with GR effects were (partially) wrongly rotated to resemble the central emission in polarization convention. So one should revisit most of the GR discussion in \cite{Marin2018b}. The GR effects play almost no role in the type-2 viewing angle, not even in the polarization angle, which was claimed previously. In general, the effects of black-hole spin, the incident polarization state, or more precisely the coronal geometries, are largely impossible to probe by X-ray polarimetry in type-2 AGNs (although some academic discussion of minor changes to the polarization output that were highlighted in \cite{Marin2018c} still remains valid). The effects of disc ionization state can be now added to this discussion, because they manifest in the global view similarly to other model dependencies related to the inner-most regions \citep[see ][for the detailed central emission comparisons with respect to the level of disc ionization]{Podgorny2023}.

We note that further deviations from the predicted total AGN output will arise from a non-isotropic treatment of the inner illumination, which was not yet tested by us. Here throughout we only illuminated the polar and equatorial scatterers by uniform {\tt KYNSTOKES} output pre-computed for $i = 20\degr$ and $i = 70\degr$, respectively, i.e. we use a semi-isotropic approximation. We also note that in the pre-computations of the central radiation the used lamp-post model is a toy model in a sense of purely isotropic coronal emission, which we do not expect, given the more sophisticated modelling estimates by the {\tt MONK} \citep{Zhang2019b, Ursini2022} or {\tt kerrC} \citep{Krawczynski2022b} computations for lamp-post coronae. Moreover, a completely different polarization properties are expected from other types of coronae appearing in the literature \citep[see e.g.][]{Marinucci2018, Poutanen2018, Krawczynski2022b, Ursini2022}. This will play a role for the type-1 AGNs, which has been also recently observationally investigated and confirmed for particular case studies. The IXPE polarimetric analysis successfully constrained some X-ray coronal properties of particular radio-quiet unobscured accreting black holes \citep{Marinucci2022, Krawczynski2022, Gianolli2023, Tagliacozzo2023, Ingram2023}. For type-2 AGNs, we predict that the properties of incident emission are rather washed out for a distant observer by the reprocessings. However, a detailed simulation would be necessary to confirm, especially for extended coronae that would not have negligible size with respect to the circumnuclear components, which is assumed here.

We also find that the total emission is highly dependent on the choice of resulting
inclination bin, in the adopted resolution of 20 angular bins between the polar and equatorial direction equally distributed in cosines of inclination $\mu_\textrm{e} = \cos (i)$. Here we chose to display the bins of $\mu_\textrm{e} = 0.375$ and $\mu_\textrm{e} = 0.975$ for type-2 and type-1 AGNs, respectively, being close to the $i = 70\degr$ and $i = 20\degr$ viewing angles in \cite{Marin2018b,Marin2018c}, but if one chooses e.g. the neighboring simulation bins $\mu_\textrm{e} = 0.325$ and $\mu_\textrm{e} = 0.925$, $\Psi$ can change in the order of a few degrees and $p$ in the order of $1\%$. Especially the common energy transition in $p$ and $\Psi$ that often served as a diagnostic tool for subtle effects in the old publications is shifted more dramatically by the choice of the inclination bin than by any other claimed effects. Because the geometrical parameters of the parsec-scale components imprint into this viewing angle effect via scattering, absorption and radiative transfer in between various defined regions, one should first try to constrain the half-opening angle (hereby fixed at $\Theta = 60\degr$, but see Paper I. for details), the shape, relative size of the equatorial and polar scatterers, the composition of the distant components and the inclination of the observer as much as possible for a particular source that one wishes to study, using other observational data from the literature. A misalignment causing a symmetry breaking and different mutual orientation of the parsec-scale components is also plausible and expected to affect the result \citep[see e.g.][]{Goosmann2011}. Only then it is feasible to proceed to discuss subtle effects arising from the inner components, especially for type-2 AGNs. The importance of geometrical parameters was already analyzed in Paper I. for reprocessing in one component, thus this conclusion for a full self-consistent model is not surprising.

\section{Observational prospects}\label{observations}

This section provides a brief summary of testing of the AGN large-scale toy models discussed in Section \ref{fullmodel} inside the {\tt IXPEOBSSIM} observation simulation software \citep[version \textit{30.2.1},][]{Baldini2022}, designed for the IXPE mission, which is operating in the 2--8 keV band. The software includes all up-to-date instrumental response matrices (version 12), but note that as a first order approximation, we will only compare the unfolded model polarization with the minimum detectable polarization, $M\!D\!P$, which is provided by {\tt IXPEOBSSIM} alongside a simulated observation and which states the polarization fraction above which any polarization is detected at more than 99 \% confidence level \citep{Fabiani2014}. Moreover, {\tt IXPEOBSSIM} currently enables only the standard unweighted analysis for pure observation simulations \citep{Baldini2022}, although the weighted approach applied on real observations could decrease the MDP by more than 10 \% due to the overall increased sensitivity of the photoelectric polarimeter when applying the weighted method \citep{DiMarco2022}.

We aim to provide crude estimates on the detectability of the total X-ray polarization signal from both type-1 and type-2 radio-quiet AGNs without focusing on a particular object. Alongside the set of observations by IXPE that was already performed (see Section \ref{introduction}), it should also give indications on which AGN configurations are favourable towards a 2--8 keV polarization detection and possible system parameter fitting through X-ray spectro-polarimetry. Due to symmetry of the system, the model polarization angle is either parallel or perpendicular to the model axis of symmetry, which is aligned with the coordinate system on the sky of the simulated observation. In order to discuss the observed polarization angle efficiently, we will use the notation of positive or negative polarization degree $p$ throughout this section, which will correspond to the parallel or perpendicular corresponding polarization angle with respect to the model axis of symmetry, respectively.

\begin{figure}
	\includegraphics[width=1.\columnwidth]{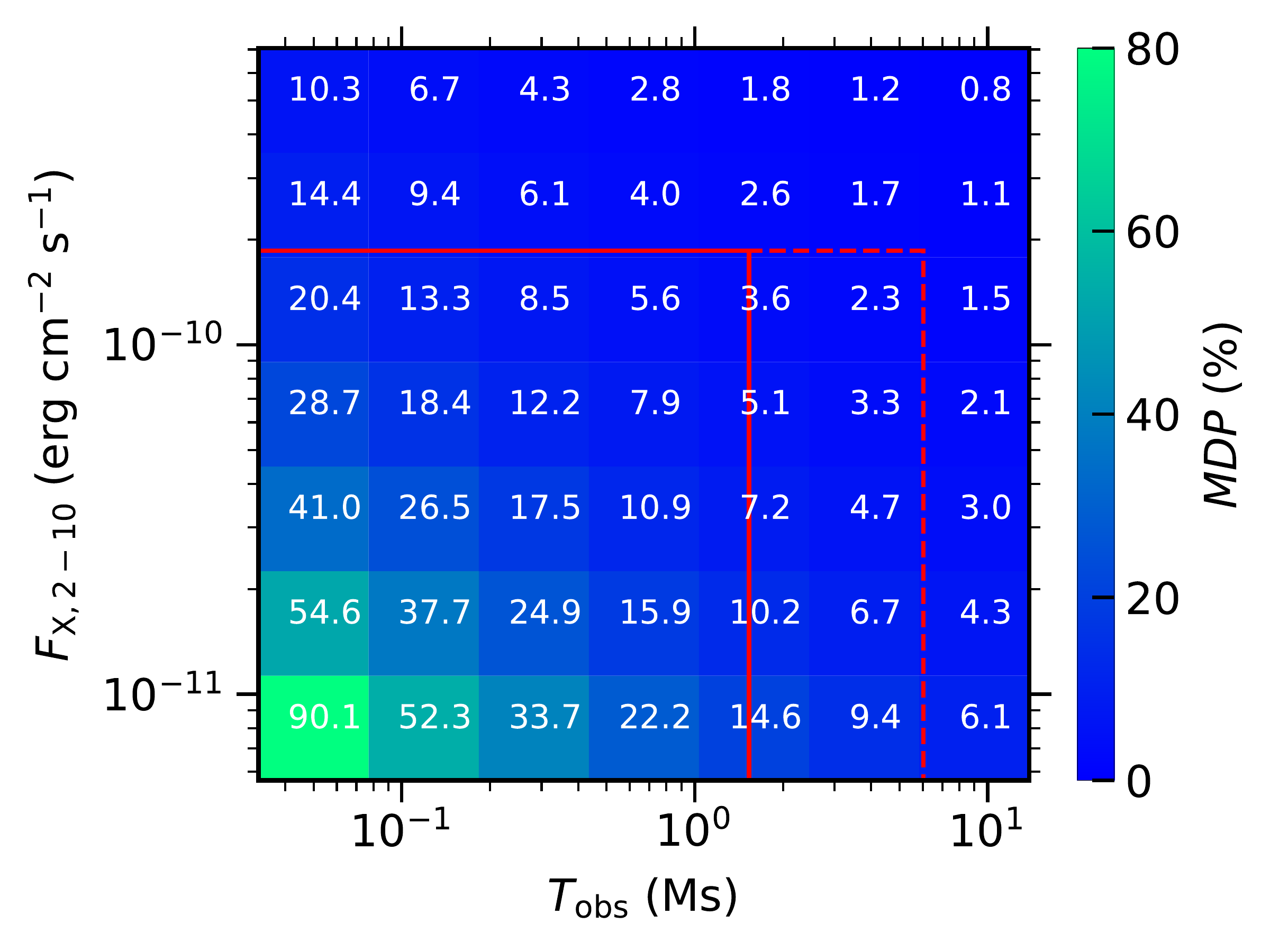}
	\caption{An example of the $M\!D\!P$ obtained with {\tt IXPEOBSSIM} in one \text{2--8 keV} energy bin for various observational times $T_\textrm{obs}$ and observed X-ray fluxes $F_\mathrm{X, 2-10}$ of a type-1 AGN model input with $2\%$ parallely polarized lamp-post emission, neutral accretion disc extending to the ISCO, black-hole spin $a = 1$, absorbing winds of $N_\textrm{H}^\textrm{wind} = 10^{21}  \textrm{ cm}^{-2}$ and equatorial torus of $N_\textrm{H} = 10^{24}  \textrm{ cm}^{-2}$. The solid red rectangle suggests a somewhat realistic window for an IXPE detection that is on one hand given by the mission's observational strategy in the point-and-stare regime and on the other hand by the brightest AGNs on the sky. The dashed line shows how the window would enlarge for eXTP in the same energy band, given its planned four times larger effective mirror area compared to IXPE.}
	\label{mdp_example}
\end{figure}

The underlying models are fully described in Sections \ref{methods} and \ref{fullmodel}. In addition to the model parameters of the unfolded spectro-polarimetric signal that serves as an input to {\tt IXPEOBSSIM}, we can adjust the galactic absorption $N_\textrm{H}^\textrm{gal}$, the single-exposure observation time $T_\mathrm{obs}$, and the X-ray source flux $F_\mathrm{X, 2-10}$ between 2 and 10 keV. We chose to display for each set of global parameter values a heatmap of the absolute value of the model polarization degree over the minimum detectable polarization, $|p|/M\!D\!P$, in the color code (below 1 is black, which means not detectable at a 99\% confidence level) versus the observed source flux $F_\mathrm{X, 2-10}$ versus the observational time $T_\mathrm{obs}$. One may assume a limit on the brightness from known stable type-1 AGN sources in the upper half of the displayed $y$-axis range \citep[$F_\mathrm{X, 2-10} \approx 1.8 \times 10^{-10} \, \textrm{erg cm}^{-2} \, \textrm{s}^{-1}$, ][]{Beckmann2006, Ingram2023}. The brightest type-2 AGNs are detected in X-ray fluxes at least an order of magnitude lower \citep[$F_\mathrm{X, 2-10} \approx 1.5 \times 10^{-11} \, \textrm{erg cm}^{-2} \, \textrm{s}^{-1}$, ][]{Bianchi2002, Tanimoto2022}. The realistic maximum observational time focused on one source reserved by IXPE is in the middle of the displayed $x$-axis range ($T_\textrm{obs} \approx 1.5 \, \textrm{Ms} \approx 17.4 \, \textrm{days}$).\footnote{Information from the IXPE team private communication, meaning the maximum observational time reserved in the mission's observing history per one source per one observing window in multiple subsequent exposures according to the telemetry limits. Note that for some stable sources the IXPE observational plan allows additive observations in multiple observational windows, such as in \cite{Tagliacozzo2023}.} However, these plots can be equally useful for predictions for the enhanced X-ray Timing and Polarimetry (eXTP) mission \citep[due to be launched in the second half of 2020s,][]{Zhang2016, Zhang2019}, operating in 2--10 keV with similar instruments on board compared to IXPE. One should then multiply the limits on the $x$-axis by a factor of $\approx 4$, by which the effective mirror area will extend for eXTP compared to the mirrors on board of IXPE \citep{Zhang2019}. Figure \ref{mdp_example} shows an example of the plain IXPE $M\!D\!P$ values in one 2--8 keV energy bin in such 2D parameter space with the suggested boundaries of the current detectability limits. In reduced energy range the $M\!D\!P$ typically increases due to the lower number of photons, while the dependency of $M\!D\!P$ on energy is more complex.\footnote{The $M\!D\!P$ not only depends on the energy-dependent number of photons $N$ as $\sim 1/\sqrt{N}$, but also on the energy-dependent modulation factor $\mu$ of the instrument (typically increasing with energy in 2--8 keV) as $\sim 1/\mu$, and on the quantum efficiency of the gas pixel detector $\epsilon$ (typically decreasing with energy in 2--8 keV) as $\sim 1/\sqrt{\epsilon}$ \citep{Fabiani2014}.}

We will again show only some representative results, but we performed the {\tt IXPEOBSSIM} simulations for the following large set of generic AGN configurations. The central lamp-post accreting models were tested for two extreme black-hole spin values $a = \{0,1\}$, highly neutral and highly ionized disc extending to the ISCO, and various values for the polarization state of the incident coronal radiation: $p_0 = \{ 0 \%, 2 \%, -2 \%\}$.\footnote{But note that the IXPE observations of the Seyfert type-1.2 AGN IC 4329A \citep{Ingram2023} and the black-hole XRB Cyg X-1 in the hard state \citep{Krawczynski2022} revealed a 2--8 keV polarization detection of ($3.3 \pm 1.9$) \% (a detection almost at the 99 \% confidence level) and ($4.0 \pm 0.2$) \% (a detection at higher than 99 \% confidence level), respectively. In both discoveries the polarization angle was consistent with the alignment of the large-scale radio jet and the polarization signatures can be attributed directly to the plasma forming the hot X-ray corona. A $\gtrsim 4$ \% 2--8 keV polarization parallel to the radio emission orientation was also attributed to the corona in the IXPE observation of the type-1.5 to type-1.8 ``Changing-look'' AGN in NGC 4151 \citep{Gianolli2023}.} The pure type-1 configurations (under the viewing angle of 20$^{\circ}$ and the half-opening angle of 60$^{\circ}$) were considered by us only for the absorbing polar winds with column densities $N_\textrm{H}^\textrm{wind} = \{10^{21}  \textrm{ cm}^{-2}, 10^{22}  \textrm{ cm}^{-2}, 10^{23}  \textrm{ cm}^{-2}\}$, and for an equatorial torus with a column density of $N_\textrm{H} = \{10^{23}  \textrm{ cm}^{-2}, 10^{24}  \textrm{ cm}^{-2}, 10^{25}  \textrm{ cm}^{-2}\}$. The pure type-2 configurations (under the viewing angle of 70$^{\circ}$ and the half-opening angle of 60$^{\circ}$) were considered for three type of polar winds: a) absorbing, b) ionized, and c) no winds, and three torus column densities $N_\textrm{H} = \{10^{23}  \textrm{ cm}^{-2}, 10^{24}  \textrm{ cm}^{-2}, 10^{25}  \textrm{ cm}^{-2}\}$. To discuss the detection limits, we used the full energy range operated by IXPE (2--8 keV) as a single energy bin. But note that in some cases, e.g. \cite{Ursini2023} for AGNs, IXPE found statistically stronger detections in restricted energy ranges, which is out of the scope of this paper to examine. We fixed the galactic absorption to $N_\textrm{H}^\textrm{gal} = 5 \times 10^{20} \, \textrm{cm}^{-2}$. In the following subsections we provide general order-of-magnitude prospects, as opposed to the handful of specific sources observed by IXPE in the first year and a half of operations that had the highest probability of detection according to the literature and the mission's selection procedures. The complete simulation grid for type-1 and type-2 AGNs is shown also from a different perspective in Tables \ref{table_model_grid_type1} and \ref{table_model_grid_type2}, which provides the model 2--8 keV polarization state alongside the interpolated exposure times where the unfolded model polarization degree was matching the obtained $M\!D\!P$ for three selected X-ray source fluxes and vice versa the interpolated X-ray source fluxes where $|p|\approx M\!D\!P$ for three selected exposure times.

\subsection{Type-1 AGNs}

\begin{figure}
		\includegraphics[width=1.\columnwidth]{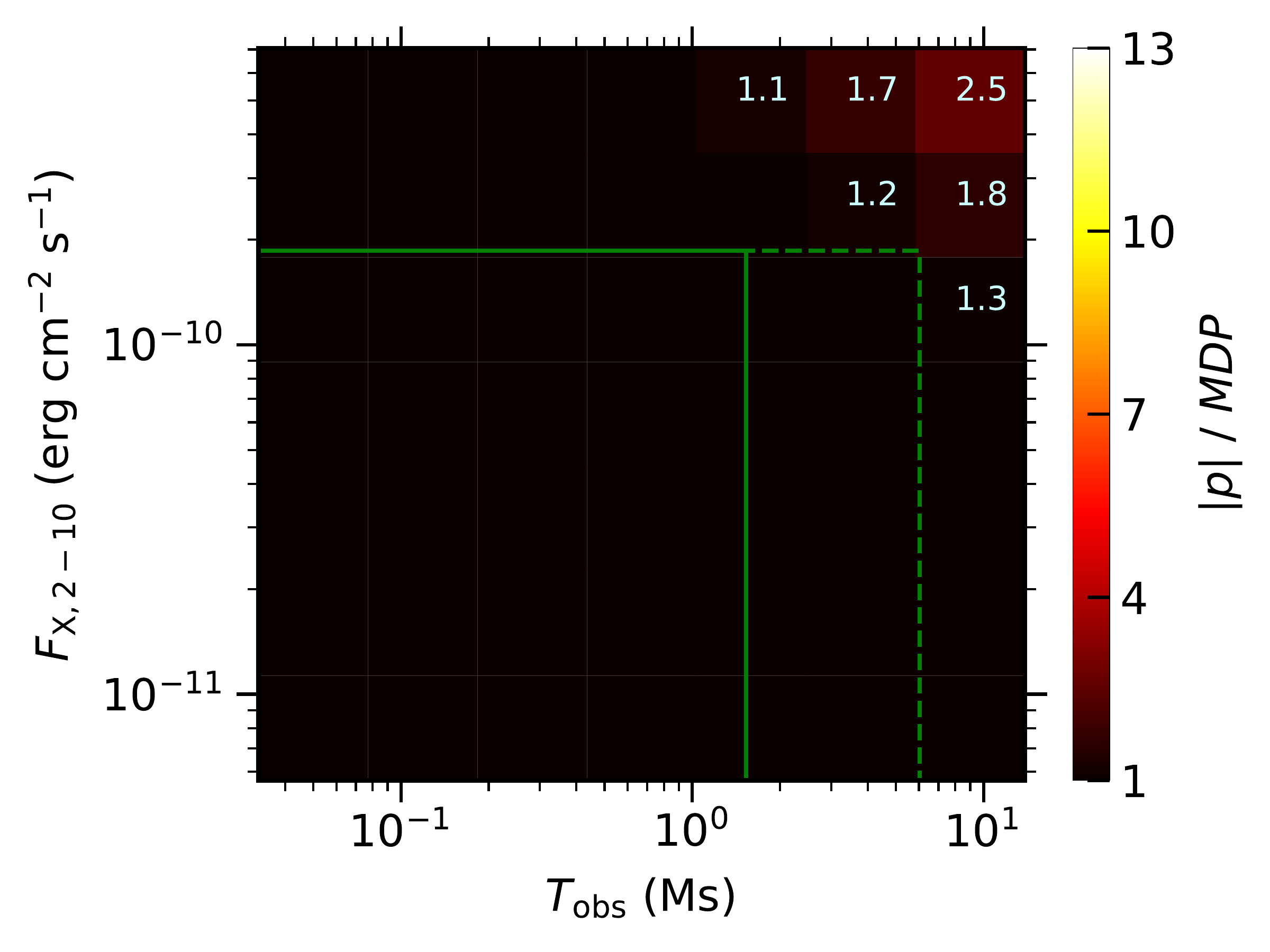}
		\caption{The model polarization degree, $p$, divided by the $M\!D\!P$ obtained with {\tt IXPEOBSSIM} in one 2--8 keV energy bin for various observational times $T_\textrm{obs}$ and observed X-ray fluxes $F_\mathrm{X, 2-10}$. We show the positive or negative sign of $p$, if the corresponding model 2--8 keV polarization angle is parallel or perpendicular to the axis of symmetry, respectively. In the color scale we show only the absolute value of $p$ divided by the $M\!D\!P$ for simplicity. If the ratio of $|p|/M\!D\!P$ is below 1, the polarization is not detected at the 99\% confidence level and we do not display the value and mark the corresponding region in black. The input model is a type-1 AGN with $2\%$ parallely polarized lamp-post emission, neutral accretion disc extending to the ISCO, black-hole spin $a = 1$, absorbing winds of $N_\textrm{H}^\textrm{wind} = 10^{21}  \textrm{ cm}^{-2}$ and equatorial torus of $N_\textrm{H} = 10^{24}  \textrm{ cm}^{-2}$, i.e. the same as in Figure \ref{mdp_example}. The solid green rectangle suggests a somewhat realistic window for an IXPE detection that is on one hand given by the mission's observational strategy in the point-and-stare regime and on the other hand by the brightest type-1 AGNs on the sky. The dashed line shows how the window would enlarge for eXTP in the same energy band, given its planned four times larger effective mirror area compared to IXPE.}
		\label{f1}
\end{figure}

The tested cases of AGNs provide in general very low chances for type-1 AGNs to be detected at the 99 \% confidence level. The $|p|/M\!D\!P$ values barely reach the ratio of 1 in the upper-right corners of the studied 2D plots. Figure \ref{f1} provides one of the possible configurations of the source: rather transparent absorbing winds, parallel oriented $2\%$ polarized primary, highly spinning black hole, and neutral accretion disc. Rest of the parameter space tested is roughly equivalent to this case or imposes even lower probability of detection. Hence there is little hope of parametric fitting with IXPE for this class of objects upon a detection, unless the source is stable and multiple observations are added \citep[as in the case of MCG 05-23-16, ][]{Tagliacozzo2023}. Although here we restrict ourselves to the unweighted analysis in {\tt IXPEOBSSIM}, which overestimates the $M\!D\!P$ values compared to the weighted analysis \citep{DiMarco2022} that was carried in majority of the IXPE discovery papers, the simulated $M\!D\!P$ values here are in line by $\sim 1$\% with those reported in e.g. \cite{Marinucci2022, Ingram2023, Tagliacozzo2023}. It is, however, fair to repeat at this point that the emission is dependent on the composition and morphology of the distant components (e.g. the half-opening angle chosen), the observer's viewing angle (that can be larger than the tested $20\degr$) and even more on the central engine model for type-1 AGNs (see Section \ref{fullmodel} for details). Thus, if we assumed higher coronal polarization than $2\%$ that the sophisticated coronal models such as {\tt MONK} or {\tt KerrC} are allowing \citep{Ursini2022, Krawczynski2022b}, the simulation results will be more optimistic. This is consistent with the statistically significant and high IXPE polarization detections attributed to the coronal power-law in \cite{Krawczynski2022, Ingram2023, Dovciak2023}. The details are related to a particular choice of coronal geometry. The probability of detection also rises, if the observer's inclination is closer to the half-opening angle, similarly to the grazing angle case of NGC 4151 \citep{Gianolli2023}. This means that a detailed case study with restricted parameter information from literature should always be simulated for particular source targeting with IXPE or eXTP. Our results should be taken as general observability prospects for AGNs.

\subsection{Type-2 AGNs}\label{type2s}

For the type-2 AGNs the situation is more favourable due to much higher polarization degree expected (tens of percents at lower energies), although the faintness of the Compton-thick sources is determinative. We display all results for a generic combination of inner-region parameters ($2\%$ parallel polarized primary, black-hole spin $a = 1$, a neutral accretion disc extending to the ISCO), as in any of the tested obscured AGNs the inner-region parameters do not affect the result (see Section \ref{fullmodel}). Because the mutual position, shape, relative size and composition of the parsec-scale scatterers is more determining for the polarization output of type-2 AGNs than of type-1 AGNs, the conclusions on detectability in the type-2 cases that we studied are also rather illustrative and one may examine even more diverse circumnuclear component configurations (see Section \ref{fullmodel} and Paper I.).

\begin{figure}
		\includegraphics[width=1.\columnwidth]{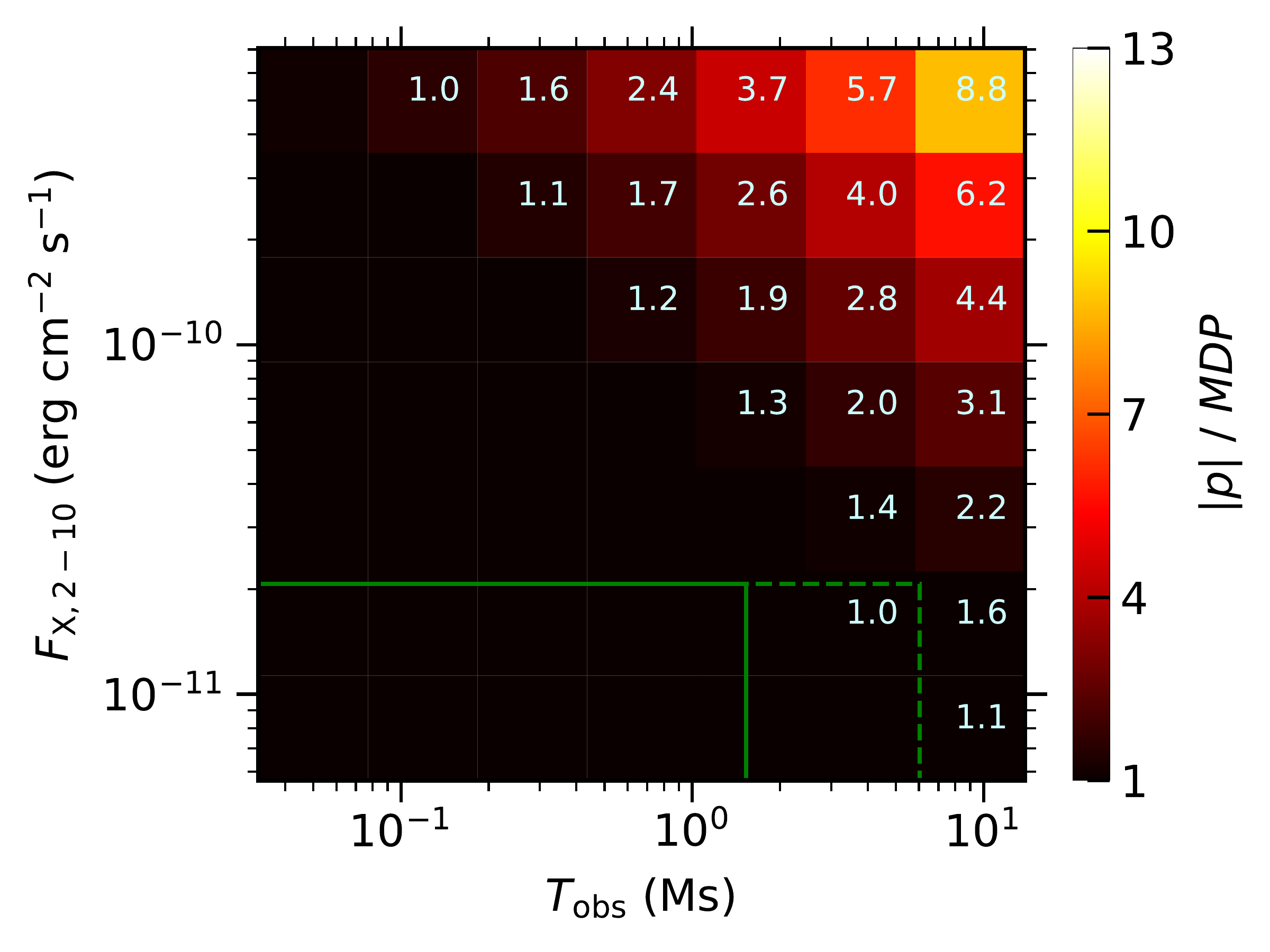}
		\caption{The same as in Figure \ref{f1}, but the input model is a type-2 AGN with $2\%$ parallely polarized lamp-post emission, neutral accretion disc extending to the ISCO, black-hole spin $a = 1$, absorbing winds of $N_\textrm{H}^\textrm{wind} = 10^{21}  \textrm{ cm}^{-2}$ and equatorial torus of $N_\textrm{H} = 10^{23}  \textrm{ cm}^{-2}$. The solid green rectangle suggests a somewhat realistic window for an IXPE detection that is on one hand given by the mission's observational strategy in the point-and-stare regime and on the other hand by the brightest type-2 AGNs on the sky. The dashed line shows how the window would enlarge for eXTP in the same energy band, given its planned four times larger effective mirror area compared to IXPE.}
		\label{f2}
\end{figure}

\begin{figure}
		\includegraphics[width=1.\columnwidth]{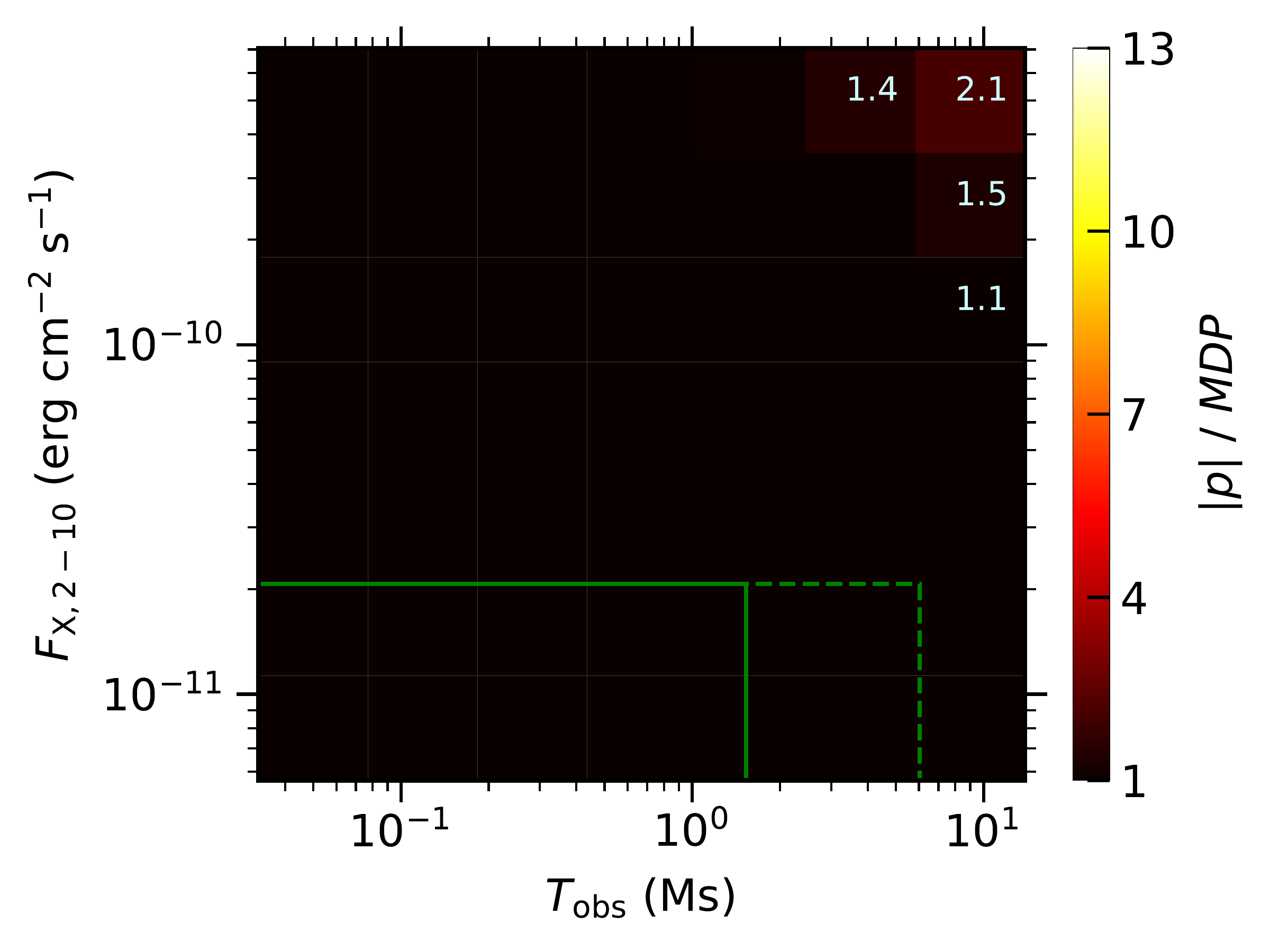}
		\caption{The same as in Figure \ref{f2}, but for torus column density \text{$N_\textrm{H} = 10^{24}  \textrm{ cm}^{-2}$}.}
		\label{f3}
\end{figure}

\begin{figure}
		\includegraphics[width=1.\columnwidth]{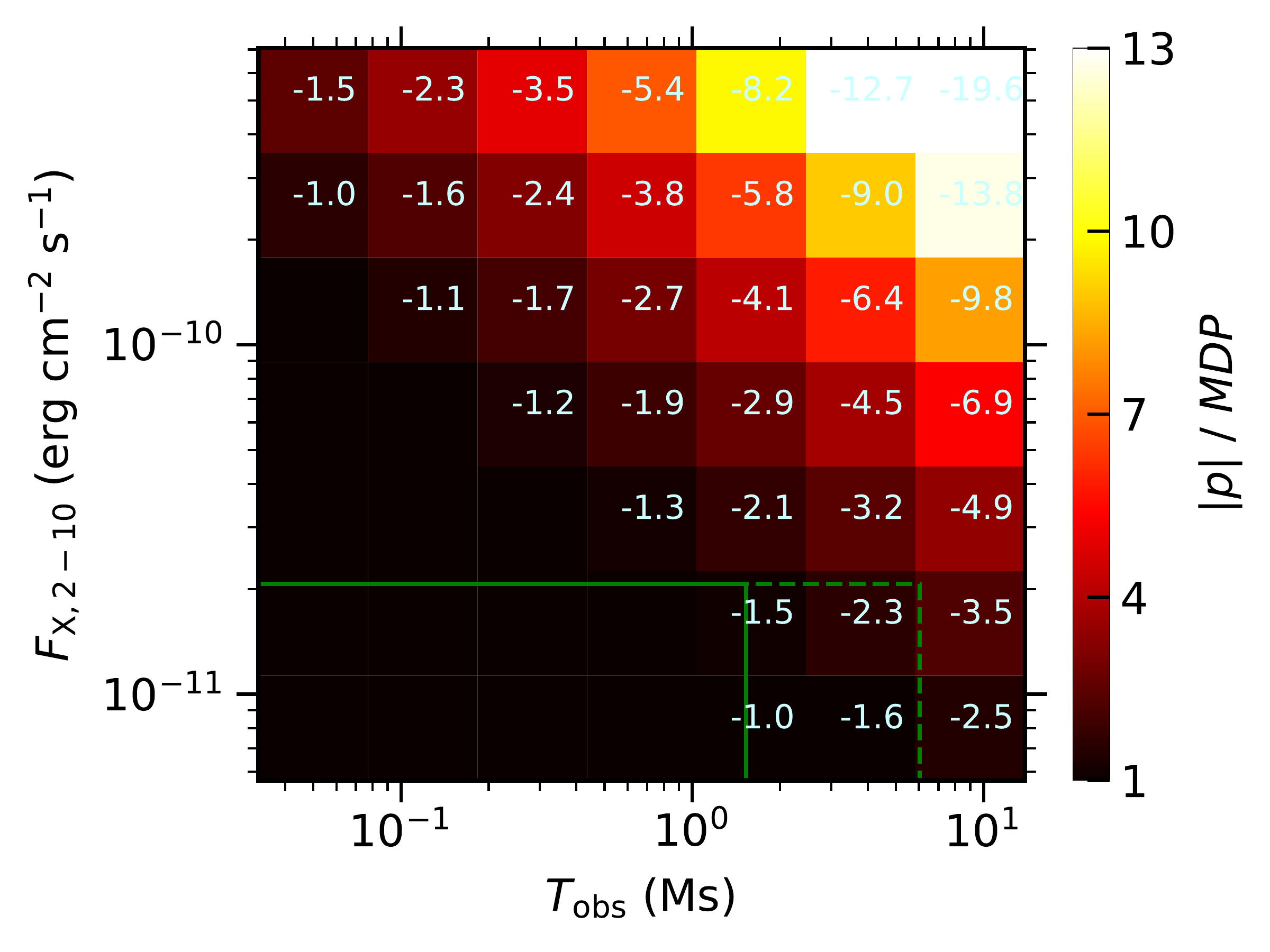}
		\caption{The same as in Figure \ref{f2}, but for torus column density \text{$N_\textrm{H} = 10^{25}  \textrm{ cm}^{-2}$}. The negative sign in front of the written values in the heatmap indicates that in this case, the net model polarization angle in 2--8 keV is orthogonal to the projected system axis of symmetry. In the color code we keep the absolute value of polarization degree for consistency of polarization detectability estimates with other cases.}
		\label{f4}
\end{figure}

Let us first consider the case of \textit{absorbing polar winds}. Figures \ref{f2}, \ref{f3}, \ref{f4} represent the cases of equatorial region column densities $N_\textrm{H} = 10^{23}  \textrm{ cm}^{-2}$, $10^{24}  \textrm{ cm}^{-2}$, $10^{25}  \textrm{ cm}^{-2}$, respectively. The more we increase the torus optical thickness, the higher polarization we expect in the total 2--8 keV band due to increase of polarization towards higher energies in the IXPE band. However, due to obscuration, the flux (and the polarized flux) have reverse dependency with torus optical thickness and energy. This trade-off regarding detectability, which applies also to the comparison between type-1 and type-2 AGNs, is articulated by the fact that should the detection be accepted at 99\% confidence level, it has to be higher than the $M\!D\!P$, which depends on the number of photons $N$ per energy band as $\sim 1/\sqrt{N}$ \citep{Fabiani2014}. It turns out that if we examine the case of torus column densities $N_\textrm{H} = 10^{23}  \textrm{ cm}^{-2}$, we simulate detections at 99\% confidence level for lower observed source fluxes and shorter exposures (Figure \ref{f2}) than for torus column densities $N_\textrm{H} = 10^{24}  \textrm{ cm}^{-2}$ (Figure \ref{f3}), which in turn provide lower detection probabilities than the cases of torus column densities $N_\textrm{H} = 10^{25}  \textrm{ cm}^{-2}$ (Figure \ref{f4}), being the most favourable configuration for a detection. In order to explain this behavior with torus column density in the 2--8 keV band, we will analyze the other cases of polar winds.

\begin{figure}
		\includegraphics[width=1.\columnwidth]{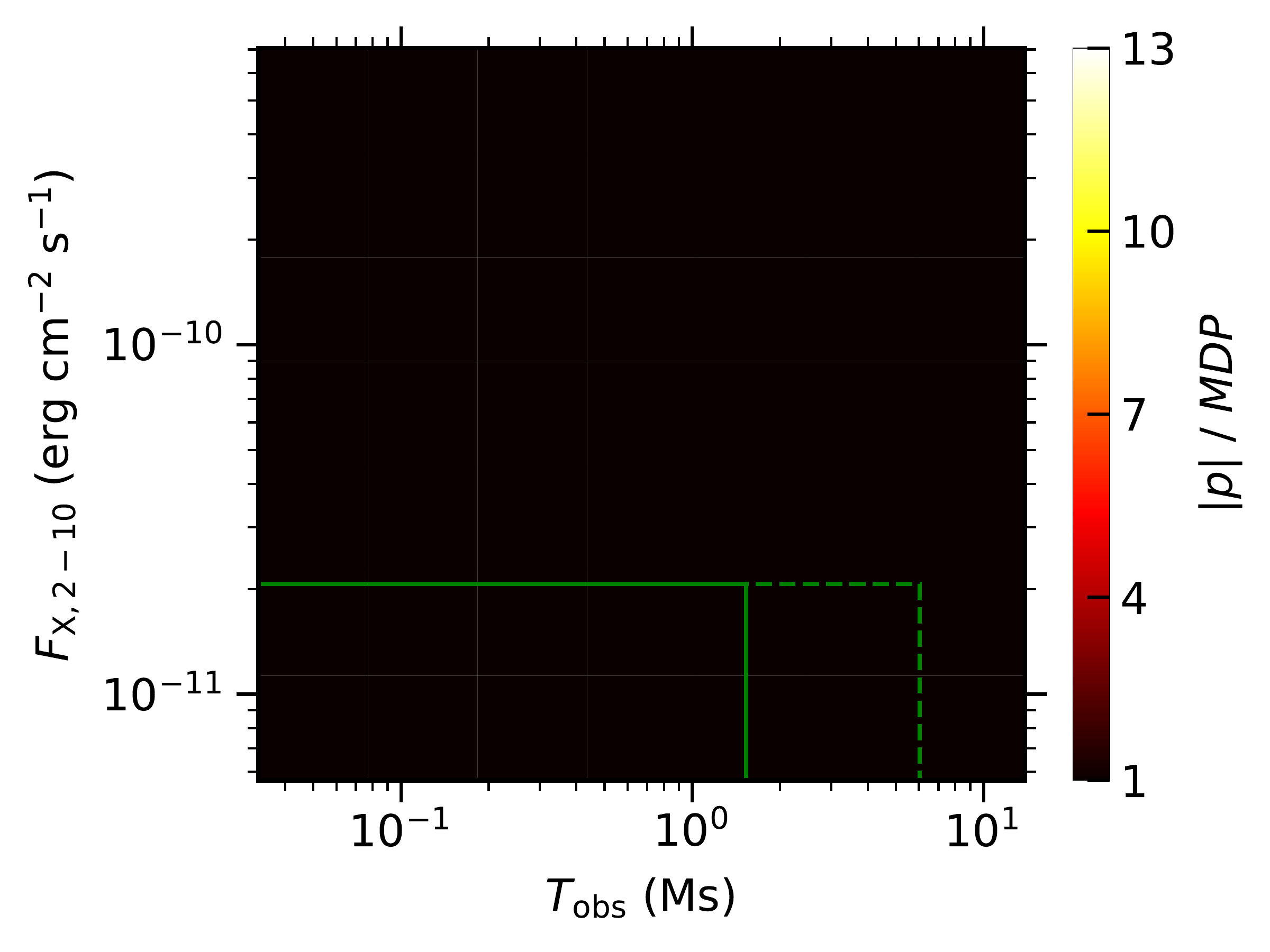}
		\caption{The same as in Figure \ref{f2}, but for ionized polar winds.}
		\label{f5}
\end{figure}

\begin{figure}
		\includegraphics[width=1.\columnwidth]{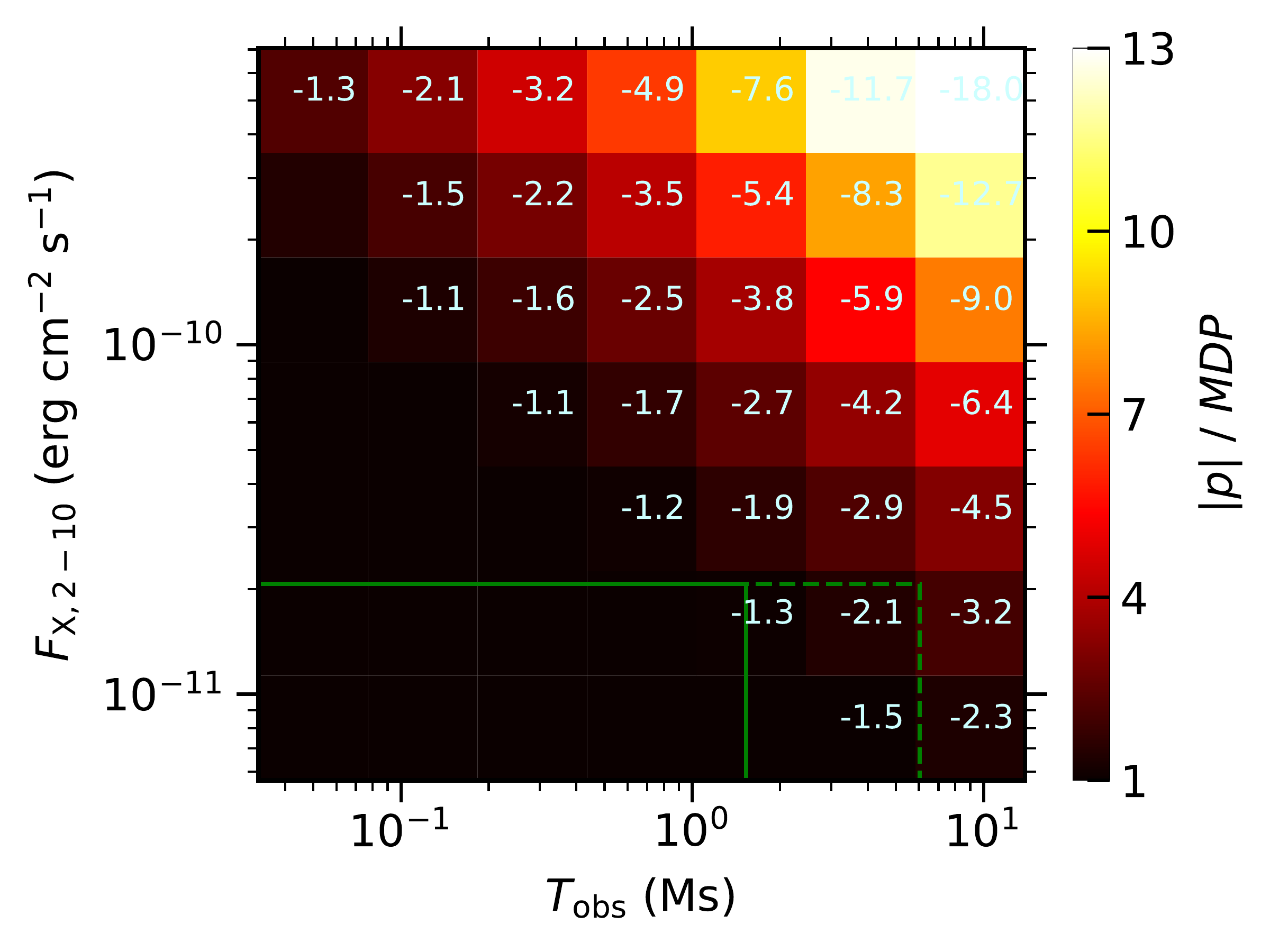}
		\caption{The same as in Figure \ref{f5}, but for torus column density \text{$N_\textrm{H} = 10^{24}  \textrm{ cm}^{-2}$}. The negative sign in front of the written values in the heatmap indicates that in this case, the net model polarization angle in 2--8 keV is orthogonal to the projected system axis of symmetry. In the color code we keep the absolute value of polarization degree for consistency of polarization detectability estimates with other cases.}
		\label{f6}
\end{figure}

\begin{figure}
		\includegraphics[width=1.\columnwidth]{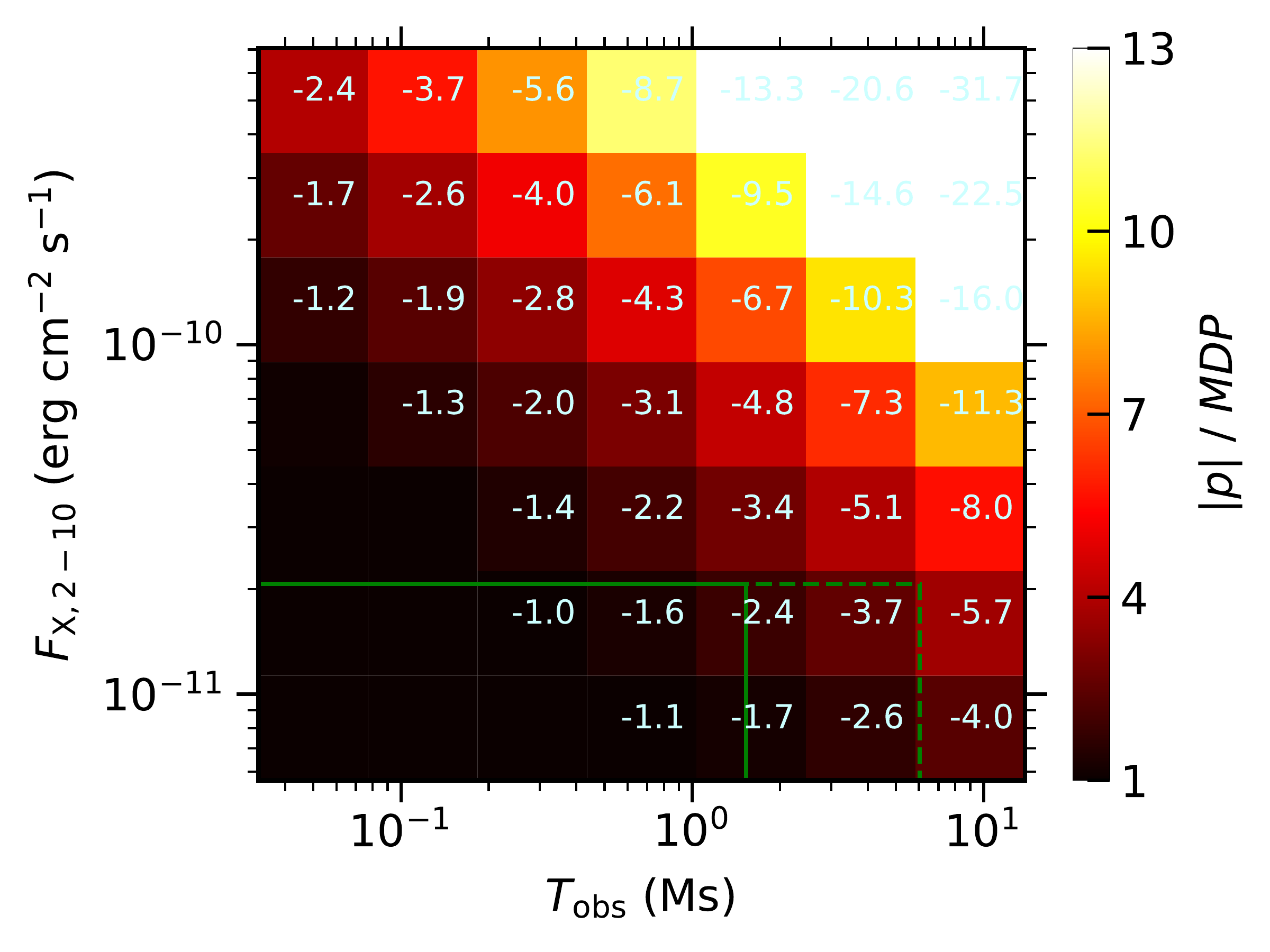}
		\caption{The same as in Figure \ref{f6}, but for torus column density \text{$N_\textrm{H} = 10^{25}  \textrm{ cm}^{-2}$}. The negative sign in front of the written values in the heatmap indicates that in this case, the net model polarization angle in 2--8 keV is orthogonal to the projected system axis of symmetry. In the color code we keep the absolute value of polarization degree for consistency of polarization detectability estimates with other cases.}
		\label{f7}
\end{figure}

\begin{figure}
		\includegraphics[width=1.\columnwidth]{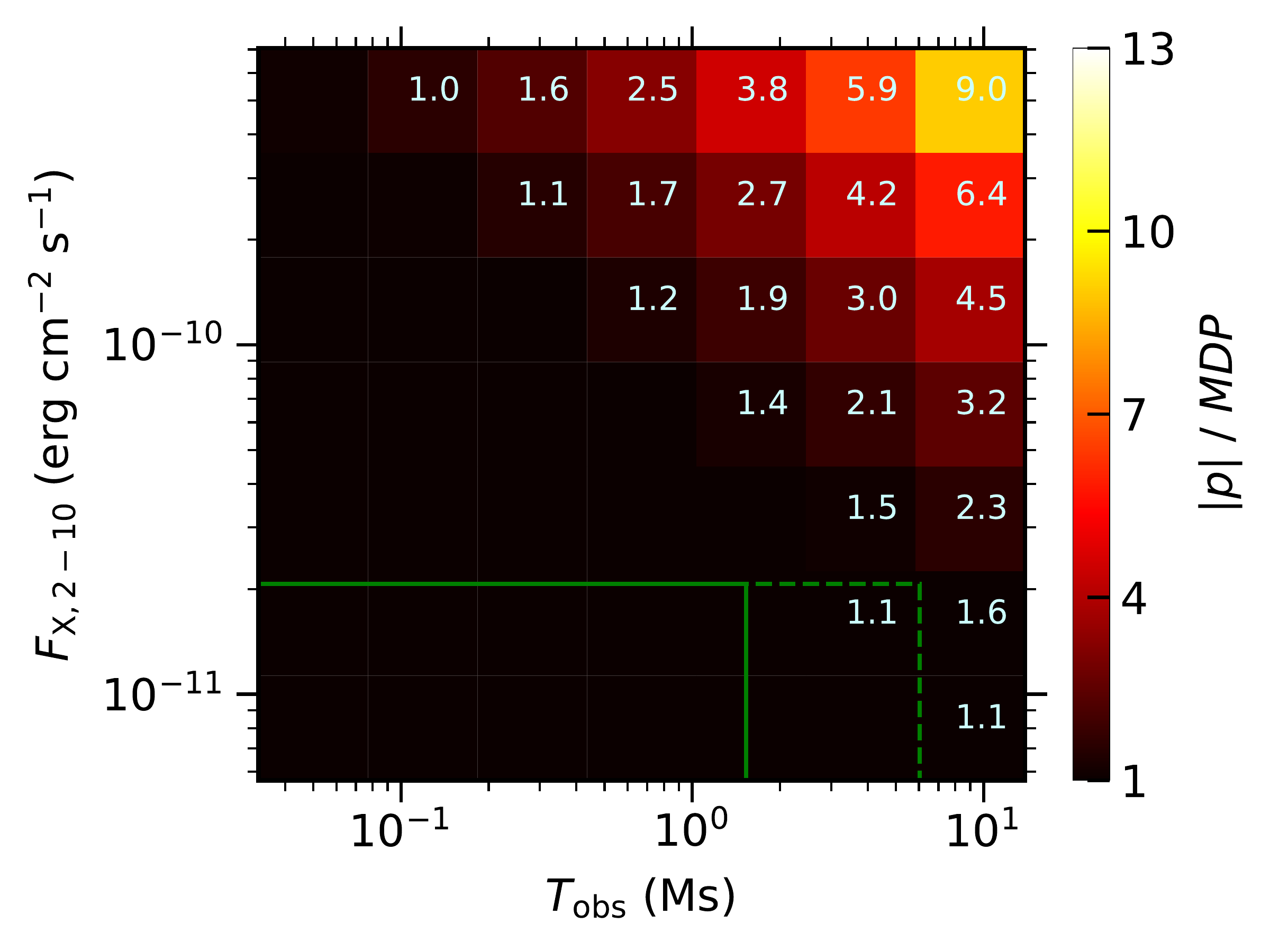}
		\caption{The same as in Figure \ref{f2}, but for no polar winds.}
		\label{f8}
\end{figure}

\begin{figure}
		\includegraphics[width=1.\columnwidth]{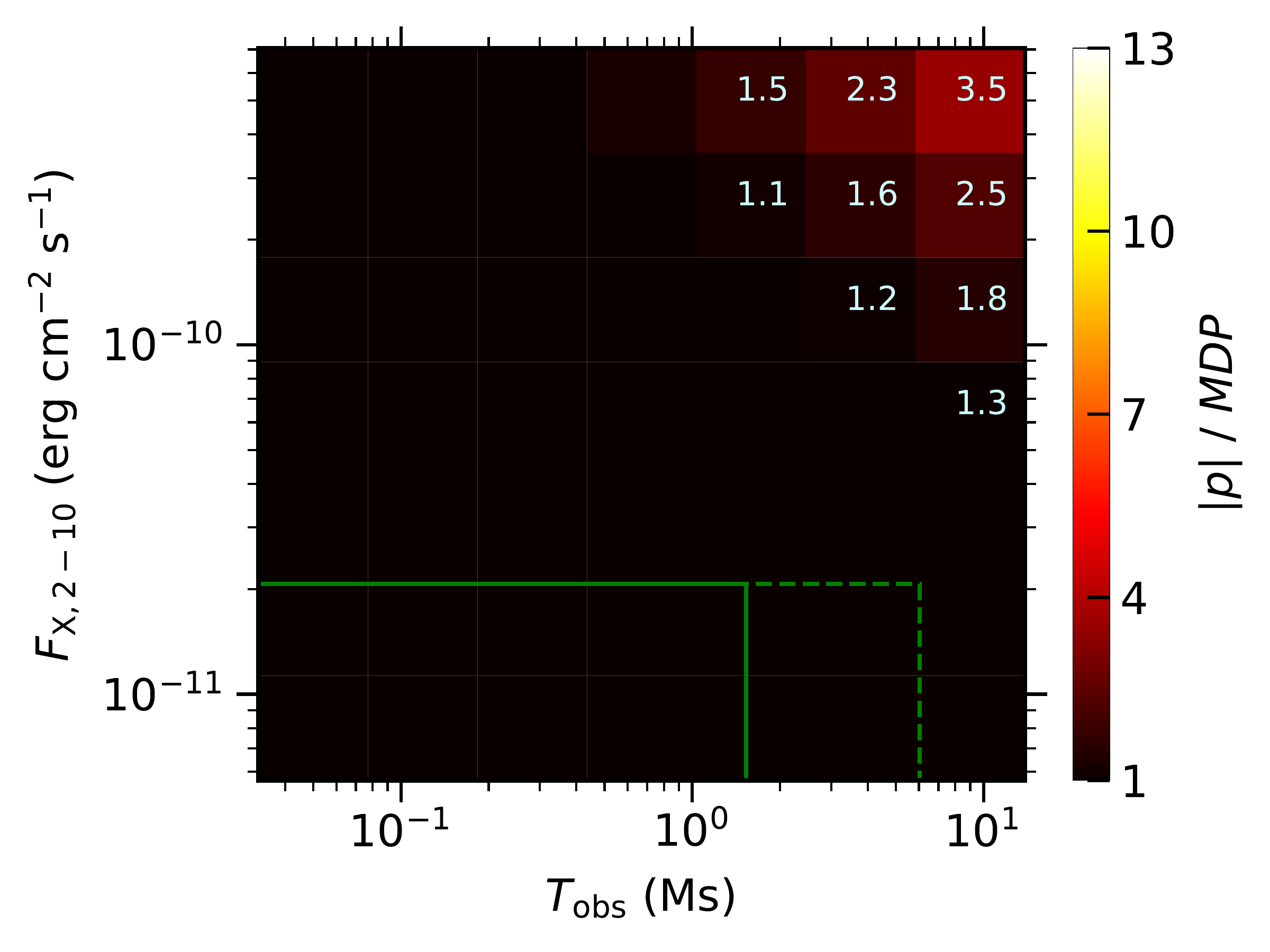}
		\caption{The same as in Figure \ref{f8}, but for torus column density \text{$N_\textrm{H} = 10^{24}  \textrm{ cm}^{-2}$}.}
		\label{f9}
\end{figure}

\begin{figure}
		\includegraphics[width=1.\columnwidth]{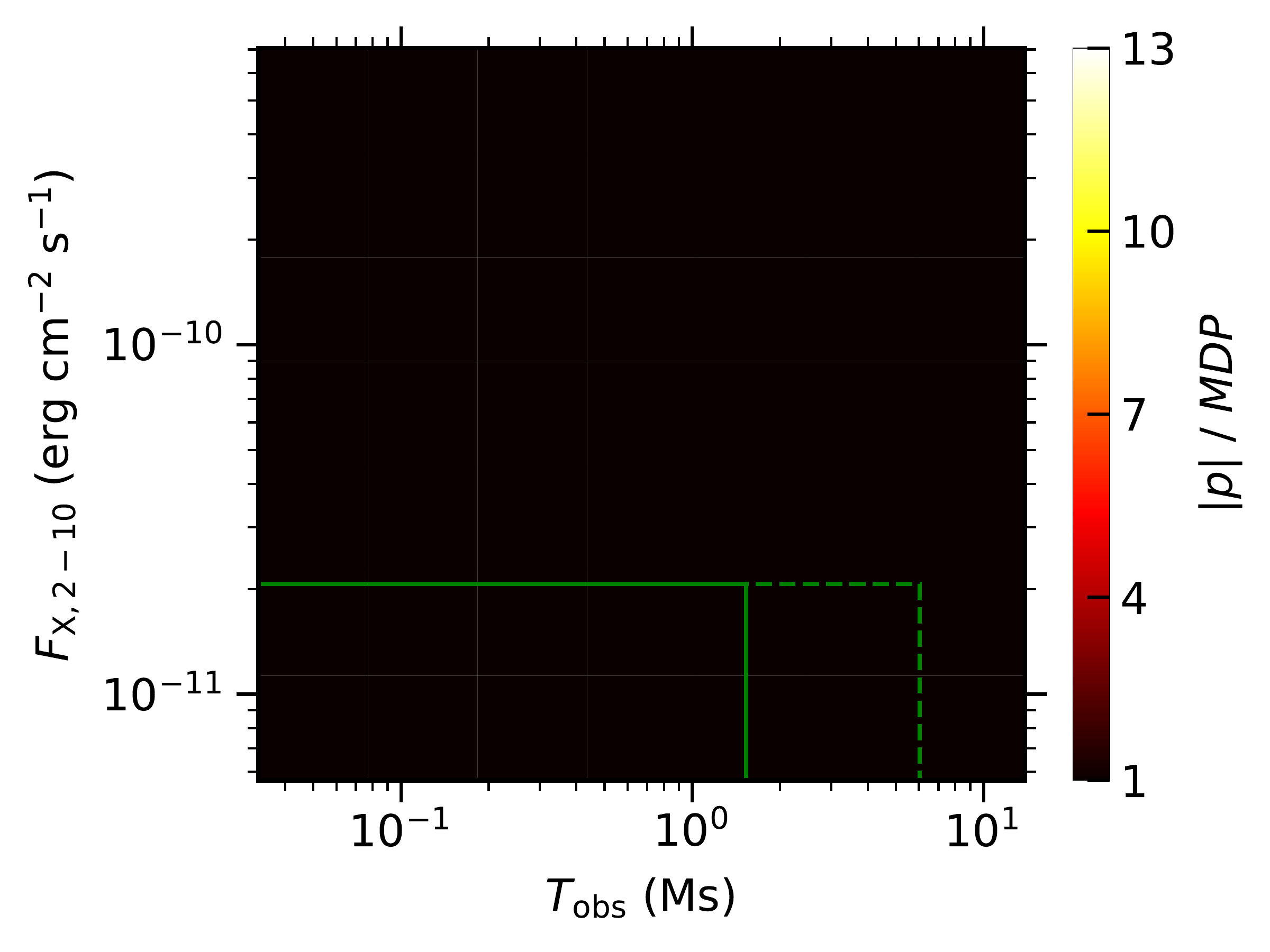}
		\caption{The same as in Figure \ref{f8}, but for torus column density \text{$N_\textrm{H} = 10^{25}  \textrm{ cm}^{-2}$}.}
		\label{f10}
\end{figure}

The composition of the polar scatterer that largely contributes to polarization at lower energies does play a role. Figures \ref{f5}, \ref{f6}, \ref{f7} represent the cases of equatorial region column densities $N_\textrm{H} = 10^{23}  \textrm{ cm}^{-2}$, $10^{24}  \textrm{ cm}^{-2}$, $10^{25}  \textrm{ cm}^{-2}$, respectively, for \textit{ionized polar winds}. We get low detection probabilities for the torus column densities of $N_\textrm{H} = 10^{23}  \textrm{ cm}^{-2}$, while significantly higher for $N_\textrm{H} = 10^{24}  \textrm{ cm}^{-2}$ and even more for $N_\textrm{H} = 10^{25}  \textrm{ cm}^{-2}$. Moreover for the lowest transparency of the torus, the detectability increases for ionized polar components compared to the absorbing polar components given the same flux and exposure time, which is intuitive as the polar material is more reflective and causes high ($\gtrsim 25\%$ perpendicularly oriented) polarization. If we do not include any polar scatterer, such as in Figures \ref{f8}, \ref{f9}, \ref{f10} representing the cases of torus column densities $N_\textrm{H} = 10^{23}  \textrm{ cm}^{-2}$, $10^{24}  \textrm{ cm}^{-2}$, $10^{25}  \textrm{ cm}^{-2}$, respectively, we get a reverse dependency of the detectability at 99\% confidence level with the torus column density, compared to the case of ionized polar winds. Thus, if there is \textit{no polar component}, the torus itself produces higher likelihood of detection if less dense. The opposite is true, if we accentuate the contribution to total polarization of the polar winds via high ionization of this component, because these are essentially reflected X-rays on the axis of symmetry (gaining high polarization through scattering at nearly $90\degr$ scattering angles) that subsequently pass through this equatorial region towards the observer. The case of absorbing polar winds represents an intermediate case. This explanation is clear from the observed polarization angle in 2--8 keV, which is in the cases of dominant polar reflection orthogonal to the axis of symmetry, i.e. orthogonal to the main plane of scattering causing the polarization. It is parallel to the axis of symmetry for the case of equatorial scattering only, as the energy transition in the polarization angle occurs rather at lower energies compared to the 2--8 keV average and taking into account the energy-dependent flux \citep[see Section \ref{fullmodel} and ][]{Marin2018b}. For the absorbing winds, if the torus column density is high enough and the polar reflection is dominant, we see a perpendicular orientation. If the torus is rather transparent, we see an average parallel orientation of polarization in 2--8 keV.

This discussion is interesting with respect to the first, and so far only, IXPE observation of a type-2 AGN in the Circinus Galaxy \citep[with Compton thickness $N_\textrm{H} > 10^{24}  \textrm{ cm}^{-2}$,][]{Arevalo2014, Kayal2023} described and interpreted in \cite{Ursini2023}. The spectro-polarimetric analysis performed in the study suggests that the observed polarization of ($17.6 \pm 3.2$) \% in the 2--8 keV band (at 68\% confidence level) can be mostly attributed to the equatorial scattering, while the polarization from the polar reflection is unconstrained.
We point to the fact that this spectro-polarimetric analysis assumes two distinct spectral power-law indices $\Gamma = 1.6$ and $\Gamma = 3.0$ for the cold (equatorial) and warm (polar) reflectors, respectively, fixed in the spectro-polarimetric fit, which results in low total flux contribution of the warm reflector, thus low contribution to the net polarization. If different assumptions were taken more in favour of the warm reflector \citep[in the appendix of ][it is discussed that such trials did not improve the spectral fit performed before the spectro-polarimetric fit and that the presented results are consistent with spectral analysis of \cite{Marinucci2013}]{Ursini2023}, it could affect the result of the polarization component analysis.
In \cite{Ursini2023} the interpretation of the observation is supported by a Monte Carlo simulation of the equatorial scatterer only, while the simulations presented in this paper and in \cite{Marin2018b} consider simultaneously the polar reflector. Because of the detectability prospects presented in this study, we humbly propose re-consideration of the contribution to the polarization detected in the Circinus Galaxy by IXPE from the partly ionized polar reflector. This is further supported by the fact that the simulations presented here and in \cite{Marin2018b} clearly distinguish the two contributions via a $90\degr$ switch in the polarization angle. Because the polarization observed in the Circinus Galaxy is orthogonal to the main axis of symmetry \citep{Ursini2023}, this would be consistent with an origin in the warm polar reflector. While Paper I. more thoroughly suggests that the parallely oriented polarization is more likely from the contribution of the pure equatorial scatterer in the 2--8 keV band, \textit{if the half-opening and the observer's inclination are high enough} \citep{Kayal2023}, assessing various geometries, column densities, and ionizations of the torus and various cases of the coronal irradiation. Although the simulations of equatorial reprocessing in the dusty torus presented in Paper I. allow a $\sim 20$\% polarization outcome perpendicular to the axis of symmetry in some configurations, such high polarization oriented perpendicularly can also plausibly arise from scattering off the NLRs and in such proposed scenario, the equatorial component may rather serve as a depolarizer to the highly polarized warm reflection, especially if polarized parallely to the principal axis.

The discussion is also interesting with respect to the energy-dependence in the 2--8 keV band. While examining various combinations of energy binning is beyond the scope of this study, we note that the contribution to polarization from the polar scatterer is typically significant at soft X-rays, extending more into the 2--8 keV range for higher opacities of the torus. The non-detection of polarization in 6--8 keV in the Circinus Galaxy is claimed in \cite{Ursini2023} to be due to the presence of unpolarized iron line, which is of course a valid point. Although the simulations presented here take into account the iron line, they cannot cover the full complexity of the line formation and its true contribution to depolarization. However, we also propose the cause of lower 6--8 keV polarization of about 10\% at 68\% confidence level \citep{Ursini2023} to be by the competing contributions of the polar and equatorial scatterers resulting in mutually orthogonal polarization vectors. For high enough column densities of the equatorial scatterer that the source possesses \citep{Arevalo2014, Kayal2023} this would not mean a switch of the polarization angle to parallel orientation at higher energies within the 2--8 keV band, but at even higher energies due to the reduced contribution from light passing directly through the opaque torus \citep[see Section \ref{fullmodel} and ][]{Marin2018b}. This is consistent with the observed polarization angle and its energy dependence \citep{Ursini2023}.

\subsection{No polar and no equatorial scatterer}\label{ixpekynstokes}

We also briefly tested the pure {\tt KYNSTOKES} lamp-post emission in {\tt IXPEOBSSIM} without the parsec-scale AGN components, i.e. the bare nucleus. Only a very small fraction of the parametric configurations covered by the {\tt KYNSTOKES} models can produce detectable polarization by IXPE at the 99\% confidence level. If the detected polarization was attributed directly to such regions comprising a toy-model primary plus relativistic reflection, the inclination of the accretion disc would have to be close to $i = 60^{\circ}$, and it would have to hold either a highly ionized disc (i.e. with high coronal luminosity and/or low black-hole mass) or a highly polarized primary (with $p_0 \gtrsim 3 \%$). This holds for nearly any realistic central black-hole spins and lamp-post heights, which are however unlikely to be examined through present-day X-ray polarimetry even in the most favourable scenarios of the lamp-post geometries. Moreover, all publications analyzing the IXPE observations of accreting black holes (both supermassive and stellar-mass) until now either preferred the interpretations of coronae extending in the accretion disc plane rather than along the principal axis (if the primary source was not obscured or if there was not only an upper limit of polarization in the X-rays), or left the question of coronal geometry unanswered.

\section{Conclusions}\label{conclusions}

Albeit a complete self-consistent X-ray polarization AGN model is a difficult task, the Monte Carlo {\tt STOKES} code allows us to study various parsec-scale scattering regions combined in 3D structures that are motivated by observational and theoretical constraints. We adopted the 3-component axially symmetric scenario introduced in \cite{Marin2018b,Marin2018c} that includes a wedge-shaped equatorial dusty torus glued to a conical polar scatterer with a central lamp-post disc-corona emission provided by the latest version of the {\tt KYNSTOKES} code. The improvements consist of partial ionization for the reflection off the upper layers of the accretion disc, correcting the radiative transfer of coronal emission and fixing the {\tt STOKES} simulation setup, including a unified notation for polarization quantities. The new computations were compared to the previous results from \cite{Marin2018b,Marin2018c}. We confirm the basic prediction for a significantly high $\gtrsim 20$\% polarization perpendicular to the axis of symmetry in the soft X-rays for type-2 AGNs due to scattering in the polar regions and the drop of polarization to $0 \% \lesssim p \lesssim 10 \%$ in the hard X-rays with a $90$\degr switch in the polarization angle due to partial transparency of the torus for type-2 AGNs. We also confirm the basic results for type-1 AGNs, i.e. that in general a polarization of up to $\sim 5$\% is expected in the entire 1--100 keV band with energy dependence and polarization angle dependent on the central primary source of emission.
 
The characteristics of the inner-most accretion regions cannot be probed by contemporary X-ray polarimetry for Compton-thick AGNs, including the GR effects in the vicinity of the central black hole, which we do not see in our simulations. Such properties are observable for type-1 AGNs, although a careful consideration of the parsec-scale equatorial and polar components is necessary, as it impacts the emission by changes of the order of $\sim 1\%$ in polarization fraction. For type-1 AGNs we predict the polarization of $\sim 1$--$2\%$ in 10--100 keV, which is about half of the previously estimated value in \cite{Marin2018c}. Our model is limited in the approximation of central emission and semi-isotropic illumination of the parsec-scale components. We also point out the importance of the observer's inclination and the mutual orientation, shape, relative size and structure of the parsec-scale components for the reprocessed radiation, especially for the X-ray polarimetry of type-2 AGNs. Although we have tested only a limited part of the feasible self-consistent scenarios for AGNs in this paper, Paper I. already revealed the full diversity of the distant equatorial reprocessings in AGNs. We conclude that albeit X-ray polarimetry is by itself already a useful ``microscope'' on the Compton-thick AGNs, its power truly emerges with each and every information from other observational techniques, which lift the anticipated degeneracies that were quantitatively illustrated in Paper I. and here. Although the first AGN observations by IXPE proved that for the \textit{brightest} and \textit{well-known} sources the X-ray polarization data can enhance our knowledge significantly, the complexity and diversity of AGNs is preventing us from providing an efficient diagnostic tool or a qualitative guidance for a \textit{general} set of sources, given the contemporary sensitivity of X-ray polarimetric instruments. Such goals are ambitious for the near future.

We focused also on the problem of AGN faintness, compared e.g. to the field of Galactic accreting black holes, where the photon-demanding X-ray polarimetry has objectively higher informative potential. The entire computed model parametric space was analyzed in the integrated 2--8 keV band by {\tt IXPEOBSSIM} that produced simulated observations for various single-exposure times and observed X-ray fluxes, using the latest instrumental response matrices of IXPE. For typical observed brightness of type-1 AGNs, IXPE will have difficulties to provide statistically significant polarization detections, hence to infer more information from the signal. Because if only upper limits are obtained \citep[see e.g.][]{Marinucci2022, Tagliacozzo2023}, one is allowed to rule out only the most extreme scenarios of emission, as the conceivable system configuration space gets significantly more degenerate towards 0\% polarization. The situation is more promising for type-2 AGNs, according to our simulations. This was illustrated on the case of Circiunus Galaxy that was already observed by IXPE and where we suggest a possible alternative interpretation of the data analyzed in \cite{Ursini2023}. We also stress the improved capacities of forthcoming X-ray polarimeters, such as the eXTP mission that will reduce the needed observational times by a factor of $\approx 4$. From the models presented in Section \ref{fullmodel} we deduce the importance of energy-dependent polarization analysis in the X-ray band, which can be further probed -- in combination with IXPE or eXTP -- by hard X-ray polarimeters, such as the XL-Calibur 15--80 keV balloon experiment \citep{Abarr2021}, or soft X-ray polarimeters, such as the REDSoX 0.2--0.8 keV sounding rocket mission \citep{Marshall2018}.

\section*{Acknowledgements}

JP acknowledges financial support from the Charles University, project GA UK No. 174121, and from the Barrande Fellowship Programme of the Czech and French governments. MD thanks for the support from the Czech Science Foundation project GACR 21-06825X. JP and MD thank also for the institutional support from the Astronomical Institute RVO:67985815.

\section*{Data Availability}

The {\tt KYNSTOKES} and its documentation is available at \url{https://projects.asu.cas.cz/dovciak/kynstokes}. The {\tt STOKES} code version \textit{v2.07} is currently not publicly available and it will be shared upon reasonable request. The analysis and simulation software {\tt IXPEOBSSIM} developed by the IXPE collaboration and its documentation is available publicly through the web-page \url{https://ixpeobssim.readthedocs.io/en/latest}.



\bibliographystyle{mnras}
\bibliography{example} 




\appendix

\section{Figures comparing the input radiation of the full AGN model}\label{total_comparisons}

In this section we display the figures related to the discussion in Section \ref{fullmodel}.

\begin{figure}
	\includegraphics[width=1.\columnwidth]{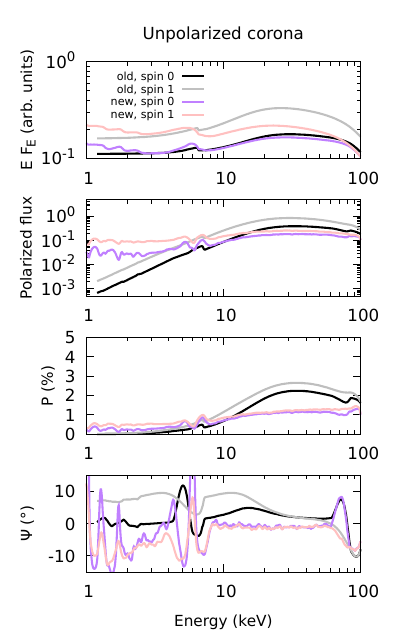}
	\caption{The incident radiation in the polar directions for type-1 AGNs in the case of unpolarized coronal radiation. We display from top to bottom the energy-dependent flux $EF_\textrm{E}$ (in arbitrary units), the polarized flux, the polarization degree and the polarization angle. The computations from \citet{Marin2018c} are displayed in black and gray for black-hole spin 0 and 1 respectively. The new computations for \textit{ionized} disc are displayed in purple and pink for black-hole spin 0 and 1 respectively.}
	\label{ionizedKY_TF_PO_PA_inputs}
\end{figure}
\begin{figure}
	\includegraphics[width=1.\columnwidth]{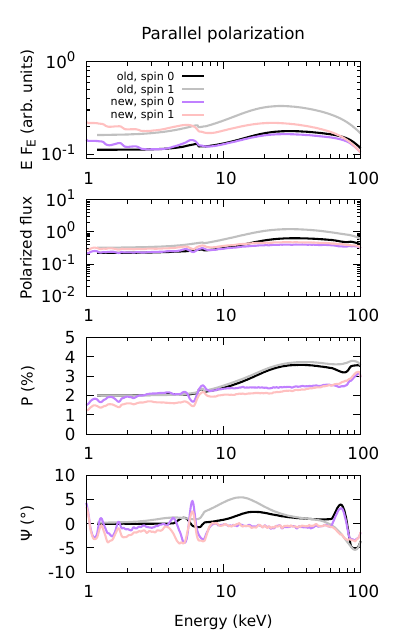}
	\caption{The same as Figure \ref{ionizedKY_TF_PO_PA_inputs}, but for $2\%$ parallelly polarized coronal radiation.}
	\label{ionizedKY_PARA_TF_PO_PA_inputs_20deg}
\end{figure}
\begin{figure}
	\includegraphics[width=1.\columnwidth]{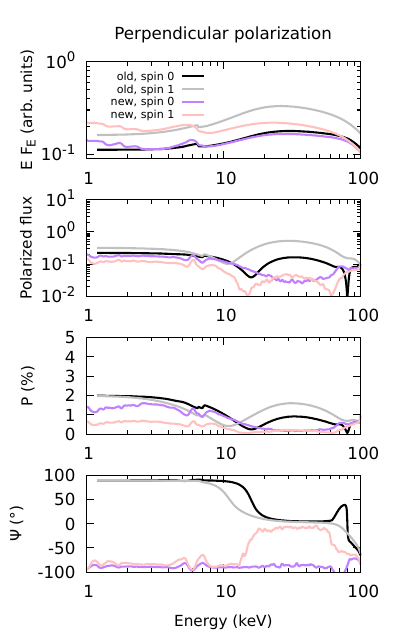}
	\caption{The same as Figure \ref{ionizedKY_TF_PO_PA_inputs}, but for $2\%$ perpendicularly polarized coronal radiation.}
	\label{ionizedKY_PERP_TF_PO_PA_inputs_20deg}
\end{figure}

\begin{figure}
	\includegraphics[width=1.\columnwidth]{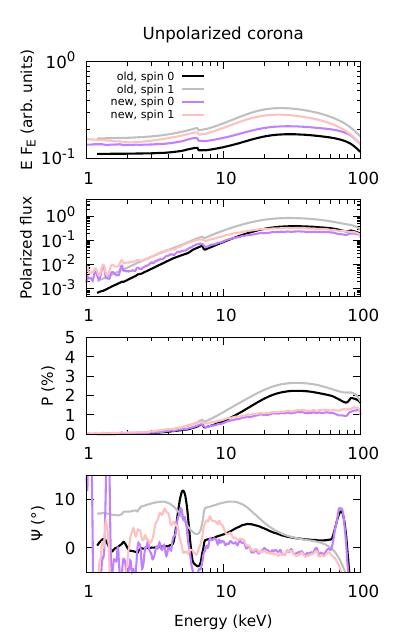}
	\caption{The same as Figure \ref{ionizedKY_TF_PO_PA_inputs}, but for \textit{neutral} disc.}
	\label{neutralKY_TF_PO_PA_inputs}
\end{figure}
\begin{figure}
	\includegraphics[width=1.\columnwidth]{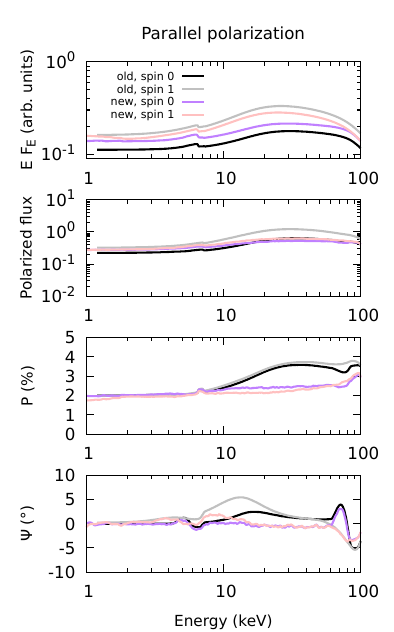}
	\caption{The same as Figure \ref{neutralKY_TF_PO_PA_inputs}, but for $2\%$ parallelly polarized coronal radiation.}
	\label{neutralKY_PARA_TF_PO_PA_inputs_20deg}
\end{figure}
\begin{figure}
	\includegraphics[width=1.\columnwidth]{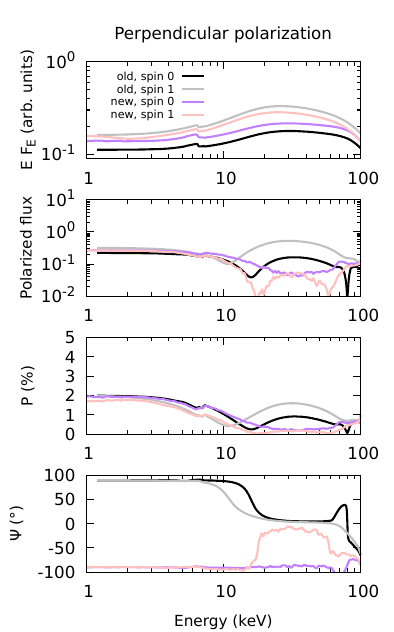}
	\caption{The same as Figure \ref{neutralKY_TF_PO_PA_inputs}, but for $2\%$ perpendicularly polarized coronal radiation.}
	\label{neutralKY_PERP_TF_PO_PA_inputs_20deg}
\end{figure}
\begin{figure}
	\includegraphics[width=1.\columnwidth]{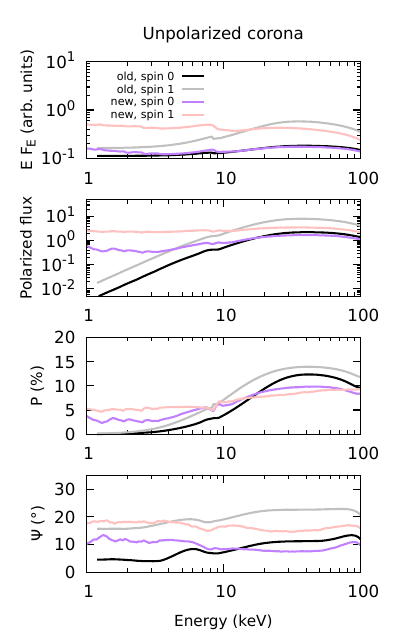}
	\caption{The incident radiation in the equatorial directions for type-2 AGNs in the case of unpolarized coronal radiation. We display from top to bottom the energy-dependent flux $EF_\textrm{E}$ (in arbitrary units), the polarized flux, the polarization degree and the polarization angle. The computations from \citet{Marin2018b} are displayed in black and gray for black-hole spin 0 and 1 respectively. The new computations for \textit{ionized} disc are displayed in purple and pink for black-hole spin 0 and 1 respectively.}
	\label{type2_ionizedKY_TF_PO_PA_inputs}
\end{figure}
\begin{figure}
	\includegraphics[width=1.\columnwidth]{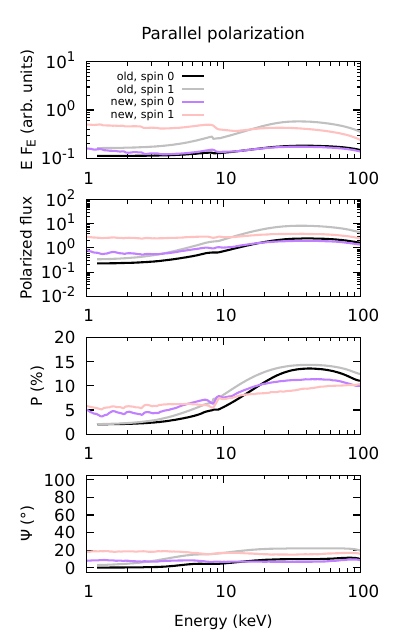}
	\caption{The same as Figure \ref{type2_ionizedKY_TF_PO_PA_inputs}, but for $2\%$ parallelly polarized coronal radiation.}
	\label{type2_ionizedKY_PARA_TF_PO_PA_inputs_70deg}
\end{figure}
\begin{figure}
	\includegraphics[width=1.\columnwidth]{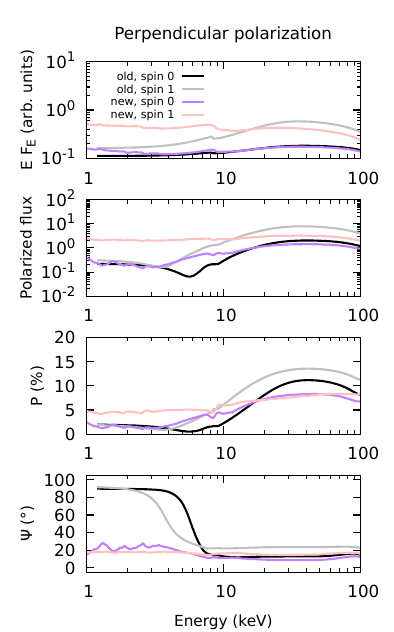}
	\caption{The same as Figure \ref{type2_ionizedKY_TF_PO_PA_inputs}, but for $2\%$ perpendicularly polarized coronal radiation.}
	\label{type2_ionizedKY_PERP_TF_PO_PA_inputs_70deg}
\end{figure}

\begin{figure}
	\includegraphics[width=1.\columnwidth]{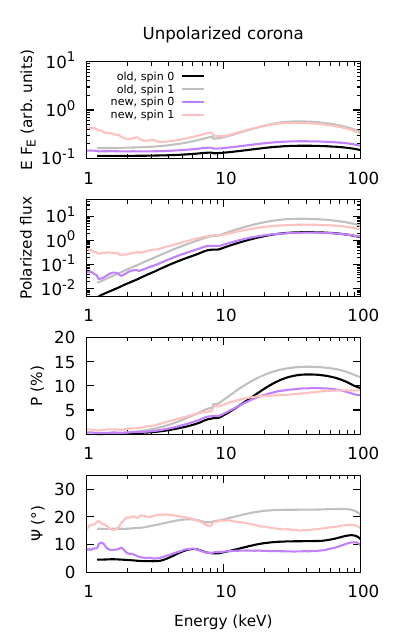}
	\caption{The same as Figure \ref{type2_ionizedKY_TF_PO_PA_inputs}, but for \textit{neutral} disc.}
	\label{type2_neutralKY_TF_PO_PA_inputs}
\end{figure}
\begin{figure}
	\includegraphics[width=1.\columnwidth]{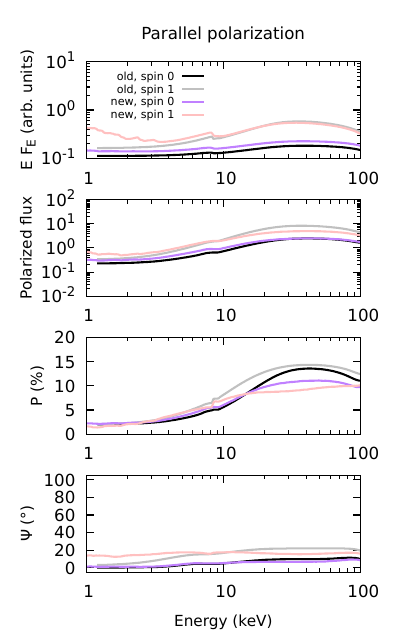}
	\caption{The same as Figure \ref{type2_neutralKY_TF_PO_PA_inputs}, but for $2\%$ parallelly polarized coronal radiation.}
	\label{type2_neutralKY_PARA_TF_PO_PA_inputs_70deg}
\end{figure}
\begin{figure}
	\includegraphics[width=1.\columnwidth]{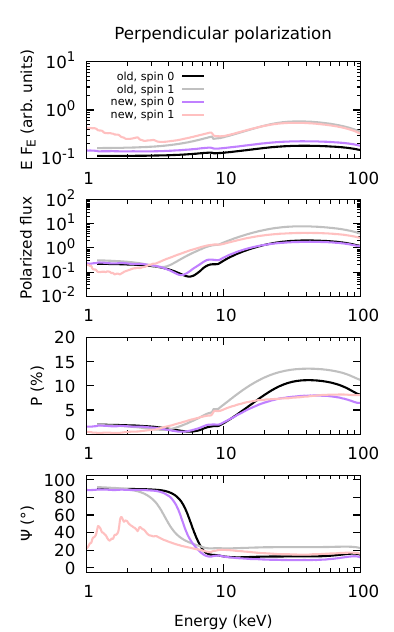}
	\caption{The same as Figure \ref{type2_neutralKY_TF_PO_PA_inputs}, but for $2\%$ perpendicularly polarized coronal radiation.}
	\label{type2_neutralKY_PERP_TF_PO_PA_inputs_70deg}
\end{figure}


\bsp	
\label{lastpage}
\end{document}